\RequirePackage{fix-cm}
\documentclass[smallextended,natbib]{svjour3}

\usepackage{amsmath,amssymb,amsfonts}
\usepackage{algorithmic}
\usepackage{graphicx}
\usepackage{textcomp}
\usepackage{xcolor}
\usepackage{array}
\usepackage{makecell}
\usepackage{url}
\usepackage{comment}
\usepackage[shortlabels]{enumitem}
\usepackage{multirow}
\usepackage{tabularx}
\usepackage{calc}
\usepackage[figuresleft]{rotating}
\usepackage{tcolorbox}

\newcommand{\revision}[1]{\textcolor{black}{{#1}}}

\newcommand{\revisiontwo}[1]{\textcolor{black}{{#1}}}

\newcommand{\boxedtext}[2]{\begin{tcolorbox}[colback=gray!10, colframe=gray!50, boxrule=0.5pt,
    arc=2pt, left=6pt, right=6pt, top=4pt, bottom=4pt, title=#1]
  #2
  \end{tcolorbox}
}

\journalname{Empirical Software Engineering}

\begin{document}

\title{Identifying Quality Indicators in Student Self-Reflections in Software Engineering}

\author{Matthew Minish  \and
        Matthias Galster  \and
        Fabian Gilson 
}

\institute{Matthew Minish \at
              University of Canterbury \\
              Christchurch, New Zealand \\
              \email{matthew.minish@pg.canterbury.ac.nz}
           \and
           Matthias Galster \at
              University of Bamberg \\
              Bamberg, Germany \\
              University of Canterbury \\
              Christchurch, New Zealand \\
              \email{mgalster@ieee.org}
           \and
           Fabian Gilson \at
              University of Canterbury \\
              Christchurch, New Zealand \\
              \email{fabian.gilson@canterbury.ac.nz}
}

\date{Received: date / Accepted: date}

\maketitle

\begin{abstract}
\textbf{Context:} 
\revision{Reflection is a fundamental skill in software engineering education, particularly in project-based courses where students learn through extended group work and need to develop their ability to reflect iteratively throughout their work. For students to benefit from reflection, their written reflections need to be assessed so that feedback can guide and improve their reflective practice.}
However, manually assessing written reflections to guide reflections is time-consuming. 
\revision{Furthermore, assessment of reflections often results in broad, non-specific feedback for a student to improve.}

\textbf{Objective:} 
\revision{This study builds on reflective writing frameworks to produce an eight-indicator scheme for assessing student reflections in software engineering.}
Furthermore, this study validates an automated classifier for assessing reflections against the framework, enabling scalable and structured feedback whilst reducing \revision{instructor} workload.

\textbf{Method:} We adapted existing reflection frameworks through iterative refinement to create our eight-indicator framework. Three annotators labelled student reflection texts, establishing moderate to reliable inter-rater agreement. We then trained and evaluated multiple encoder-only transformer models and compared them with decoder-only large language models using zero-shot prompting.

\textbf{Results:} The fine-tuned RoBERTa model achieved the strongest performance, substantially outperforming decoder-only models in both accuracy and speed. The classifier demonstrated human-level agreement on most indicators whilst enabling near-instantaneous classification. We provide two model variants optimised for different assessment priorities.

\textbf{Conclusions:} 
Our fine-tuned encoder-only models enable efficient automated assessment of reflective writing. 
\revision{The framework and automated classifier offer a means to provide timely, structured feedback on student reflections in software engineering.}

\end{abstract}

\section{Introduction}
\label{sec:introduction}

\subsection{Problem and Motivation}

In software engineering, reflecting on one's own behaviour is a widely used and established element of professional learning and continuous improvement \citep{babbEmbeddingReflectionLearning2014,guoReviewProjectbasedLearning2020,rogersReflectionHigherEducation2001}. Through reflection, software engineers leverage experiences, knowledge, and emotions to gain further insights for future situations \citep{moonHandbookReflectiveExperiential2004}. In educational settings, reflections are often part of experiential learning such as software engineering project courses \citep{morales-trujilloThreeYearStudyPeer2022}. 
In industry, reflections are essential for continuous improvement and adapting how practitioners work \citep{babbEmbeddingReflectionLearning2014}. Indeed, reflection is not only one of the 12 principles of the agile software development \citep{beckManifestoAgileSoftware2001}, but is also made explicit through everyday practices such as the daily stand-up meeting \citep{schwaberScrumGuide2011} and retrospectives \citep{derbyAgileRetrospectivesMaking2006}. \cite{schonReflectivePractitionerHow1983} captures this in the concept of the `reflective practitioner'.  
Reflections cover various aspects, such as reviewing the outcomes of individuals' (or teams') processes and practices, but also considering the effects of behaviours or skills, and how these may be altered for different future outcomes \citep{moonHandbookReflectiveExperiential2004,zimmermanBecomingSelfRegulatedLearner2002}.

Reflection itself is a skill that can be trained \citep{moonHandbookReflectiveExperiential2004}. Indeed, teaching \textit{how} to reflect, reason, and develop problem-solving skills is a key outcome in software engineering education and training \citep{turnsIntegratingReflectionEngineering2014,chngIncorporatingReflectionComputing2018}. 
\revision{While reflection can occur in many forms---such as through group discussions or reflection workshops---written reflections are most common in project-based software engineering education \citep{shekarActiveLearningReflection2007,hazzanReflectionAbstractionLearning2005,priorReflectionHardTeaching2016}. These typically take the form of journaling or formal reflection reports, though they may also be produced through specialised digital tools such as Microsoft Viva\revisiontwo{~\citep{microsoftMicrosoftVivaInsights2026}} or GitHub's Good Day Project~\citep{kalliamvakouOctoverseSpotlight20212021}, or through survey-based approaches~\citep{meyerEnablingGoodWork2021}.}
To help students improve the quality of their reflection in such writing, it is important to \emph{assess} reflections and give feedback on reflections \citep{acm/ieeeACMIEEEJoint2014}. 
\revision{However, such assessment is time-consuming, and the feedback tends to be broad and non-specific~\citep{poldnerAssessingStudentTeachers2014}---for instance, commenting on the overall quality of a reflection without identifying which particular dimensions (such as reasoning, consideration of others' perspectives, or future planning) a student has addressed or neglected. More targeted feedback that identifies specific dimensions of reflection to improve has been shown to better support the development of reflection skills~\citep{ullmannAutomatedAnalysisReflection2019}.}
\revision{More analytical approaches---where specific dimensions of a student's reflection are identified and fed back individually---can support the development of reflection skills by directing students' attention to dimensions they may have overlooked~\citep{wulffComputerBasedClassificationPreservice2021, ullmannAutomatedAnalysisReflection2019}. For example, rather than receiving feedback such as ``your reflection could be deeper,'' a student might instead be prompted to consider the perspectives of their team members, or to articulate concrete reasons for why an experience unfolded as it did. Such feedback requires first identifying which dimensions of reflection (e.g., perspectives, reasoning) are present or absent in a student's writing---a classification task that is the focus of this study.} 
\revision{In this study, we focus on \emph{structured written reflections} produced \emph{weekly} by students during a software engineering project course. Students respond to three guiding prompts that ask them to consider what went well, what did not, and what they intend to do differently, reflecting on their own work, their interactions with their team, and their progress throughout the project. This format is described in detail in Section~\ref{sec:background:reflection-journaling-context}, and the data collection context is described in Section~\ref{sec:method:data-collection-context}.}

To support more analytical assessment of written reflections, a number of reflective writing assessment frameworks have been proposed (see Section~\ref{sec:background:assessment-frameworks-for-reflective-writing}), while further studies have evaluated the feasibility of different approaches for automating the assessment of written reflections (see Section~\ref{sec:background:automated-approaches-for-assessing-reflective-writing}). 
These frameworks and automated assessment approaches have been evaluated in a number of different fields outside software engineering, e.g., health / dental students \citep{jungHowHowWell2020,jungUsingTheoryinformedData2022}, pharmacy students \citep{liAreDeeperReflectors2023,liuEvaluatingMachineLearning2019} and teacher education \citep{nehybaApplicationsDeepLanguage2023,solopovaPapagAIAutomatedFeedback2023,wulffComputerBasedClassificationPreservice2021,wulffUtilizingPretrainedLanguage2023a}. 
While recent research with computer science students may be closely related to software engineering \citep{alrashidiEducatorsValidationReflective2020,alrashidiEvaluatingAutomatedAnalysis2023,chngIncorporatingReflectionComputing2018}, the typically more individualised and short-term nature of computer science education may differ from the unique context of software engineering project-based learning, such that existing assessment frameworks and automated approaches may not transfer to the collaborative, iterative, and professional practice dimensions central to software engineering reflection. In software engineering, students typically engage in extended, group-based practical work resembling real-world development contexts, where reflection occurs iteratively throughout the project lifecycle \citep{morales-trujilloThreeYearStudyPeer2022,groeneveldSoftwareEngineeringEducation2020}. In such settings, the timeliness of providing feedback is critical as students continue working after reflecting---unlike the one-off assignments or final reflections common in other contexts where delayed or no feedback may be acceptable \citep{demmanseppThinkTwiceExploring2019,subramanianInfluenceCourseDesign2020}. 
\revision{To date, no study has developed or evaluated an automated approach for classifying reflective writing specifically within a software engineering project context.}

Regarding software engineering practitioners, some works have investigated how to improve common practices involving reflection such as daily stand-up meetings \citep{strayObstaclesEfficientDaily2013,strayDailyStandUpMeetings2020} and retrospectives \citep{derbyAgileRetrospectivesMaking2006,dubinskySoftwareGovernanceUsing2012}. 
\revision{Prior studies have explored how developers benefit from self-reflection in their day-to-day work~\citep{kalliamvakouOctoverseSpotlight20212021, meyerEnablingGoodWork2021}, finding that purposeful self-reflection can increase developers' awareness of their work habits and support self-improvement.} However, these studies focus on improving the effectiveness of existing reflection practices\revision{, or understanding their effects on developers,} rather than assessing the quality of reflective outputs, which is the focus of our work\revision{, situated in an educational context}.

\revisiontwo{Reflection among professional software engineers also differs from that of students in ways that motivate a student-specific focus. Practitioners commonly reflect collaboratively through established, team-based practices such as retrospectives and daily stand-ups~\citep{derbyAgileRetrospectivesMaking2006,strayDailyStandUpMeetings2020}, where reflection is directed at improving a shared process or product, and where experienced colleagues can scaffold the activity. Students, in contrast, are still developing reflection as a skill and commonly lack depth in their reflections~\citep{limSystematicReviewOutcomes2022,priorReflectionHardTeaching2016}: they typically reflect individually and in writing, their reflections serve their own learning rather than team process improvement, and they depend on assessment and feedback to develop their reflective practice~\citep{acm/ieeeACMIEEEJoint2014}. Supporting reflection in software engineering education therefore requires approaches tailored to individual, written reflections produced by learners---the setting addressed in this study.}
 
\subsection{Paper Goals and Contributions}
\label{sec:introduction:goals-and-contributions}

\revision{This study investigates how to support the assessment of student reflections in software engineering and in particular project-based learning. Building on prior work in reflective writing assessment in other educational contexts~\citep{chngIncorporatingReflectionComputing2018, nehybaApplicationsDeepLanguage2023, alrashidiEvaluatingAutomatedAnalysis2023, ullmannAutomatedAnalysisReflection2019}, we adapt existing reflection indicators for the software engineering context and develop an automated classifier to identify indicators in reflection text.}
We address two problems: (1) The lack of consistent and relevant feedback on reflective writing, which often results in broad or vague feedback that is not actionable; (2) The effort to provide detailed, individualised assessments. 
By extending existing reflection frameworks (see Section~\ref{sec:method}) and leveraging recent advancements in natural language processing, we automate a classification of reflective writing, enabling scalable and structured feedback while reducing instructor workload.

This study makes four key contributions:

\begin{enumerate}
    \item \revision{An adaptation of existing reflective writing frameworks for software engineering project courses,} with eight indicators that capture facets of reflection (e.g., reasoning or feelings). The \revision{adapted} framework has been validated by human \revision{instructors} who used the framework to identify indicators in reflection text from software engineering students.
    \item An automated classifier built on fine-tuned transformer models to identify reflection indicators in reflection text. The classifier achieves human-level performance on reflection assessment based on the indicators in the reflective writing framework.
    \item A comparison of encoder-only models (e.g., BERT variants) versus decoder-only models (e.g., LLMs) for this task, demonstrating the trade-offs between classification accuracy and computational efficiency. This is to show the practical applicability of the automated classifier.
    \item Practical deployment guidelines for reflection classification, including two model variants optimised for different use cases based on the weight of indicators.
\end{enumerate}

The remainder of this paper is organised as follows. Section~\ref{sec:background} provides background on reflection in software engineering education and reviews related work on automated assessment of reflective writing.
\revisiontwo{Section~\ref{sec:method} provides an overview of our research method.}
\revision{In Section~\ref{sec:reflection-assessment-framework} we present our reflective writing framework with eight indicators adapted for software engineering project-based learning.}
\revisiontwo{In Section~\ref{sec:classifier-development} we describe our methodology for annotating student reflections and developing automated classifiers, including both encoder-only and decoder-only models.}
In Section~\ref{sec:results} we present the performance of our classifiers and their agreement with human annotators. Finally, in Section~\ref{sec:discussion} we discuss the implications of our findings, limitations of the work, and provide recommendations for educators and directions for future research. We conclude in Section~\ref{sec:conclusion}. \section{Background}
\label{sec:background}

\subsection{Reflection as a Skill for Software Engineers}
\label{sec:background:importance-of-reflection}

In software engineering practice, reflection is part of professional development and continuous improvement \citep{babbEmbeddingReflectionLearning2014,galsterSoftSkillsRequired2023}. The concept of the `reflective practitioner' \citep{schonReflectivePractitionerHow1983,hazzanReflectivePractitionerPerspective2002} describes how professionals in practice use reflection to learn from experience, adapt, and improve their work.

Reflection is also one of the twelve principles of the Manifesto of Agile Software Development \citep{beckManifestoAgileSoftware2001} and is embedded within agile software development through several established practices. Daily stand-up meetings \citep{schwaberScrumGuide2011,strayDailyStandUpMeetings2020} provide opportunities for teams to reflect on progress and impediments, whilst retrospectives \citep{derbyAgileRetrospectivesMaking2006} prompt teams to reflect on their processes and identify improvements.

\subsection{Reflection in Software Engineering Training}

In software engineering training, reflection is an important element of learning, e.g., in experiential learning such as software engineering project courses \citep{shekarActiveLearningReflection2007,hazzanReflectivePractitionerPerspective2002,hazzanReflectionAbstractionLearning2005}. Learning how to reflect meaningfully is  non-trivial and at a tertiary level, students commonly lack depth in their reflections \citep{limSystematicReviewOutcomes2022,menekseReflectionInformedLearningInstruction2020}. To help students improve reflection skills, it is important to assess their reflective processes to provide feedback \citep{priorReflectionHardTeaching2016,acm/ieeeACMIEEEJoint2014}.
Furthermore, improvements in reflection quality are significantly correlated with better learning gains and student performance \citep{emboRelationshipReflectionAbility2015,maeotsRelationStudentsReflection2016,chouEffectsReflectionCategory2011,menekseReflectionInformedLearningInstruction2020}.

The current understanding of reflective practice is largely built upon the work of four different theorists (\cite{deweyHowWeThink1910}, \cite{schonReflectivePractitionerHow1983}, \cite{kolbExperientialLearningExperience2014}, and \cite{mezirowTransformativeDimensionsAdult1991}). Although each has presented their own theories for reflection, their ideas align in that reflection is a process of ``intentionally making meaning of experiences in service of future action'' \citep{turnsIntegratingReflectionEngineering2014}. Despite this general consensus over what reflection is, and the processes by which it occurs, there exists no similar standard for how the quality of a student's reflection might be assessed within a software engineering context.

\subsection{Assessment Frameworks for Reflective Writing}
\label{sec:background:assessment-frameworks-for-reflective-writing}

We can consider the assessment of reflection through two lenses \citep{moonHandbookReflectiveExperiential2004}: firstly, by evaluating the effects or outcomes---typically on learning---observed as a result of the reflection (emphasizes reflection as an element of the wider learning process); secondly, by inspecting the process of reflection itself, and analysing the processes or artefacts (such as reflective writings) produced therein (supporting the development of reflection as a skill).

An example of evaluating the effects or outcomes (learning) may be: if a student is asked to keep a learning journal in which they reflect on their ability to work with others, educators may assess reflection by whether the student's teamwork skills (the learning outcome) improve, rather than by inspecting the contents of the learning journal (the process of reflection) itself. Alternatively, when assessing the processes and artefacts of reflection directly---commonly by considering reflective writing such as essays, reflection questionnaires, or learning journals---these artefacts are typically processed in relation to taxonomies for learning or cognition, such as the Structure of the Observed Learning Outcome (SOLO) taxonomy \citep{biggsEvaluatingQualityLearning2014}, or Bloom's (Revised) Taxonomy \citep{krathwohlRevisionBloomTaxonomy2002}. Additionally, some novel frameworks have been proposed for assessment of reflective writing within the context of computer science education more specifically \citep{ullmannAutomatedDetectionReflection2015,alrashidiFrameworkAssessingReflective2020}, having been adapted from established theories of reflection and learning, and designed to better support automation of assessment with machine learning approaches \citep{ullmannAutomatedAnalysisReflection2019}. Such approaches supporting automated assessment are important as they may enable real-time and scalable feedback systems for students. Real-time feedback is normally not feasible in real educational settings given that assessment of reflection artefacts by humans is time consuming, and as a result any feedback that is given may be too holistic in nature, lacking specific guidance for each student \citep{poldnerAssessingStudentTeachers2014}.

\subsection{Automated Approaches for Assessing Reflective Writing}
\label{sec:background:automated-approaches-for-assessing-reflective-writing}

Existing approaches to automatic reflection analysis can be broadly categorized into keyword-based and machine learning-based methods ~\citep{alrashidiReflectiveWritingAnalysis2020}. Keyword-based methods, such as those proposed by \cite{ullmannArchitectureAutomatedDetection2011,ullmannReflectiveWritingAnalytics2017}, identify specific keywords in the text as indicators of reflection, e.g., words like `I' and `Feel' may be indicative of a student discussing their feelings, while words like `because' and `but' might be more indicative a student highlighting difficulties they experienced. Other approaches have been proposed which use Linguistic Inquiry and Word Count (LIWC) to identify words linked to conceptual reflective statement types---e.g., affective terms like `happy' or `anxiety', cognitive terms like `understand', or, perceptual terms like `heard' \citep{linWordcountApproachAnalyze2016}.

Machine learning-based methods have also been developed which use classification algorithms trained on annotated datasets of reflection text to detect reflective indicators. \cite{alrashidiEvaluatingAutomatedAnalysis2023} constructed a classifier for their Reflective Writing Framework \citep{alrashidiFrameworkAssessingReflective2020} tailored for computer science education, applying natural language processing techniques like n-grams and part-of-speech n-grams, along with a Random Forest classifier. \cite{nehybaApplicationsDeepLanguage2023} applied multiple techniques, including both shallow classifiers such as Random Forest, and pre-trained deep learning models such as XLM-RoBERTa, for reflective writing analysis using Ullmann's assessment framework \citep{ullmannAutomatedAnalysisReflection2019}. Pre-trained deep learning models such as XLM-RoBERTa (and other BERT variants) use a transformer architecture \citep{devlinBERTPretrainingDeep2019} and are \textit{encoder-only} models, which typically need to be fine-tuned on task-specific data to perform text classification. 

Recently, \textit{decoder-only} large language models (LLMs) such as GPT and Qwen have emerged as an alternative approach to encoder-only models \citep{naveedComprehensiveOverviewLarge2025}, demonstrating strong performance across various natural language processing tasks, including text classification \citep{galkeAreWeReally2025,kostinaLargeLanguageModels2025}.

Unlike encoder-only models that require task-specific fine-tuning, decoder-only models can perform classification through prompting approaches, where the model is provided with instructions and examples within the input prompt itself \citep{brownLanguageModelsAre2020}. This zero-shot or few-shot capability offers flexibility, as models can be adapted to new tasks without gradient updates or retraining \citep{weiChainofThoughtPromptingElicits2023}. However, the computational demands of these larger models---which typically have billions of parameters---can limit their practical deployment, particularly when real-time responses are required. Recent comparisons suggest that whilst decoder-only models show promise for classification tasks, fine-tuned encoder-only models often achieve superior performance when sufficient labelled data is available for training \citep{galkeAreWeReally2025,wangSelectingBERTGPT2024}.
 \section{Overview of Research Method}
\label{sec:method}

\begin{sidewaysfigure}[htbp]
    \includegraphics[width=\linewidth]{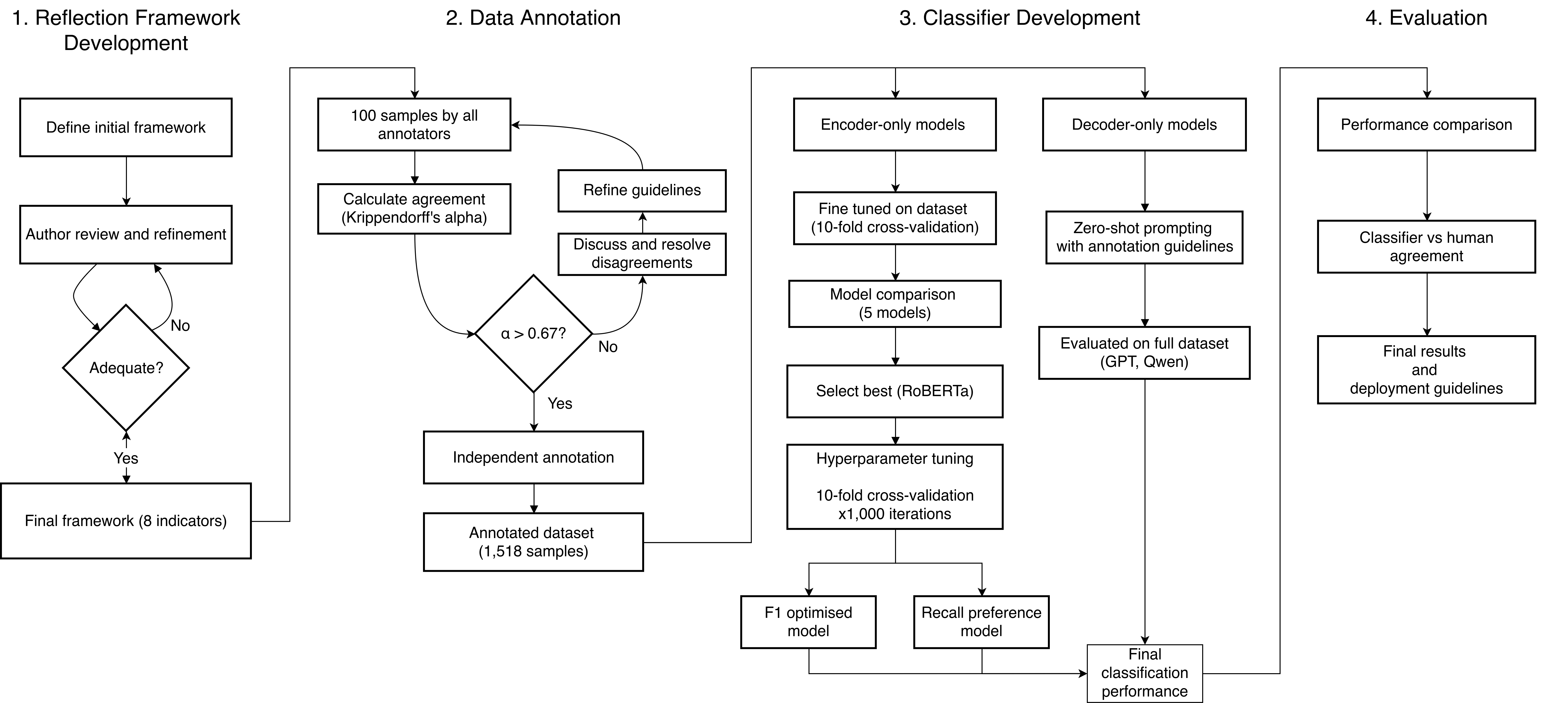}
    \caption{Overview of research method in four stages 
}
    \label{fig:method_overview}
\end{sidewaysfigure}

\revision{This section provides a brief overview of our research method, which spans Sections~\ref{sec:reflection-assessment-framework} (the reflective writing framework) and~\ref{sec:method:annotating-training-set} onwards (the annotation and classifier development). The detailed methodological descriptions follow in the subsequent sections.} 
In this study we employ a multi-stage approach to develop and evaluate an automated reflection classifier for software engineering, shown in Figure~\ref{fig:method_overview}.
 Our method comprises four main stages:
\begin{enumerate}
\item First, we developed and refined a reflective framework with eight indicators through iterative review by the authors (Section~\ref{sec:reflection-assessment-framework}). This framework was developed specifically considering the nature of reflections in software engineering projects.
\item Second, we annotated a dataset of \revision{1,518 student responses} from a software engineering context using this framework, establishing inter-rater reliability through multiple rounds of annotation and guideline refinement (Section~\ref{sec:method:annotating-training-set}). 
\item Third, we trained and evaluated multiple machine learning classifiers, comparing both encoder-only models (e.g., BERT, RoBERTa) and decoder-only models (GPT, Qwen) for automated classification (Section~\ref{sec:method:classifier}). By comparing multiple model architectures, we aimed to identify the most suitable approach for practical deployment in educational settings, where both classification accuracy and computational efficiency are important considerations.
\item Finally, we assessed the best-performing classifier's agreement with human annotators to determine whether it achieved human-level performance (Section~\ref{sec:results:classifier-vs-human-annotators}).

\end{enumerate}

The iterative nature of framework development and annotation ensured that our indicators could be applied consistently by human raters before attempting automated classification. 

\section{Reflective Writing Framework for Software Engineering}
\label{sec:reflection-assessment-framework}
\label{sec:solution:reflection-assessment-framework}

\subsection{Software Engineering Context of Reflections}

The assessment of reflections depends on the context in which reflections are done, why, and how reflections are done. This context also influences the design of the reflective writing framework.

\subsection{Context of Student Reflection}
\label{sec:background:reflection-journaling-context}

\revision{In our context, reflections follow a structured format inspired by the questions commonly used in a Scrum Retrospective~\citep{schwaberScrumGuide2011}, but are performed weekly during each sprint while the sprint work is still in progress. Unlike a Scrum retrospective, which occurs at the end of a sprint, or a final reflection submitted after a project has concluded, these reflections allow students to adapt their behaviour within the same sprint. In this sense, they function as a form of reflection-in-action~\citep{schonReflectivePractitionerHow1983}:}

\begin{enumerate}
    \item What did you do well this week?
    \item What didn't you do so well this week?
    \item What will you do differently in future?
\end{enumerate}

These questions attempt to target distinct aspects of the reflective process. In the first question, we encourage students to acknowledge their successes and consider how their practical experiences connect with theoretical knowledge—a component we view as important in project-based learning. 
\revision{Drawing on Zimmerman's~\citep{zimmermanBecomingSelfRegulatedLearner2002} account of self-regulated learning---where self-satisfaction with one's own behaviour is identified as a driver of motivation and engagement---we aim for students to consider both technical and personal / social aspects of their software development work, with the intention of fostering such feelings of self-satisfaction.}

In the second question, we encourage students to reflect on the negative aspects of their experiences, with the intention of developing students' willingness to reflect openly on failures and shortcomings. This question guides students' reactions towards being more adaptive---willing to change and learn---rather than defensive, where students might withdraw from learning opportunities.

The third question attempts to focus on future planning, where we ask students to set concrete goals and justify their chosen actions. This approach aligns with what we understand to be the `goal setting' and `strategic planning' aspects of the forethought phase of self-regulated learning \citep{zimmermanBecomingSelfRegulatedLearner2002}.

We envision reflections of varying lengths in a journaling approach, with responses ranging from a couple of brief sentences, to more comprehensive analyses over 2--3 paragraphs. 

\subsection{Design of Reflective Writing Framework}
\label{sec:method:framework-design}

The reflective writing framework (RWF) we use for assessing reflective writing of software engineering students was adapted from earlier works by \cite{ullmannComparingAutomaticallyDetected2012,ullmannAutomatedDetectionReflection2015,ullmannAutomatedAnalysisReflection2019}, and  \cite{alrashidiFrameworkAssessingReflective2020,alrashidiEvaluatingAutomatedAnalysis2023}. These frameworks were created for assessing reflective writing in computer science education \citep{alrashidiFrameworkAssessingReflective2020}, and we have adapted them to more closely suit our specific software engineering project-based context, with a reflection format constructed of three parts (see Section~\ref{sec:background:reflection-journaling-context}). \cite{ullmannComparingAutomaticallyDetected2012} initially proposed five `indicators of reflection', whose definitions have been refined and changed slightly since their initial proposal. \revision{Ullmann's original elements were: Description of an experience, Personal experience, Critical analysis, Taking perspectives into account, and Outcome~\citep{ullmannComparingAutomaticallyDetected2012}.}
More recently, Alrashidi et al. have built upon this work by proposing a new framework with seven indicators of reflection\revision{---Descriptive, Understanding, Feelings, Reasoning, Perspective,
New Learning, and Future Action---}in addition to defining three `reflection levels'---1 (non-reflective), 2 (reflective), or 3 (critically reflective)---consistent with the levels of reflection defined by Wong et al. \citep{wongAssessingLevelStudent1995}. 

\revision{In this study we adapt the indicators proposed by \cite{ullmannComparingAutomaticallyDetected2012, ullmannAutomatedDetectionReflection2015, ullmannAutomatedAnalysisReflection2019} and \cite{alrashidiFrameworkAssessingReflective2020, alrashidiEvaluatingAutomatedAnalysis2023} for the software engineering project context, with annotation criteria developed iteratively through multiple rounds of author review and refinement,}
as shown in Figure~\ref{fig:method_overview}. In Section~\ref{sec:reflection-assessment-framework:indicators-of-reflection} we describe the indicators of reflection present in this framework, including---where appropriate---how they deviate from the indicators proposed by Alrashidi et al. and Ullmann.

\revision{We note that the indicators in our framework---both in their original forms and in our adaptation---describe dimensions of reflective writing rather than properties specific to software engineering. The framework's value in our context therefore lies not in encoding software-engineering-specific concepts, but in providing a structured means to identify which facets of reflection a student has addressed, so that feedback can be targeted to facets they may have overlooked. The practical use of this targeted feedback is discussed in Section~\ref{sec:discussion:actionable-feedback}.}

\revision{While the indicators describe facets of reflective writing rather than properties specific to software engineering, they align with established themes in the software engineering literature on reflection. Critical analysis of one's actions and their consequences (Reasoning, Hindsight) is central to the ``reflective practitioner'' framing in software engineering~\citep{schonReflectivePractitionerHow1983, hazzanReflectivePractitionerPerspective2002}; reflection on perspectives within a team (Perspective) is a foundational practice in agile retrospectives~\citep{derbyAgileRetrospectivesMaking2006} and daily stand-ups~\citep{strayDailyStandUpMeetings2020}; and the formulation of intentions for future practice (Future Intention) underpins continuous improvement in agile software development~\citep{babbEmbeddingReflectionLearning2014}. The indicators thus support analysis of reflective writing that is consistent with how reflection is understood and practised in software engineering contexts.}

\subsection{Indicators of Reflection}
\label{sec:reflection-assessment-framework:indicators-of-reflection}

Each of the eight indicators described below represents a specific facet of reflection. 

The most significant change in our proposed indicators when compared to the most recent framework by \cite{alrashidiEvaluatingAutomatedAnalysis2023} is that we have broken down the `Future Action' indicator into two separate indicators. The `Future Action' in \citep{alrashidiEvaluatingAutomatedAnalysis2023} considers not just future intention by the writer---e.g., ``Next week I will...''---but also a retroactive analysis where ``the writer would, given the same circumstances again, intentionally do something differently''. In our context, students answer three questions, two of which are explicitly retrospective in nature, and the third of which is explicitly prospective. 
\revision{Beyond this structural alignment, the split reflects a conceptual distinction that is particularly important in our context: because reflections are performed weekly during an ongoing project, students have the opportunity to act on their reflections while work is still in progress. Distinguishing between retrospective analysis (how the student would have approached a past situation differently) and prospective planning (what the student intends to do next) allows us to separately identify a student's capacity for critical self-evaluation and their ability to translate that evaluation into concrete plans for action.}
For this reason, we split the `Future Action' indicator into both its forward looking component---which we call \textbf{Future \textit{Intention}} to avoid confusion with the original term---and its retrospective component---which we call \textbf{Hindsight}. Below we list the eight indicators of our framework, \revision{each with examples drawn from real student reflections}. Further examples can be found alongside annotation guidelines as supplementary material \citep{minishAssessmentSelfreflectionsSoftware2025}:

\begin{enumerate}
    \item \textbf{Description.} The writer provides a factual recounting of events they have witnessed or experienced. This does not require any degree of analysis or deeper understanding, and as such is not indicative \textit{of} reflection, but it does provide the context needed \textit{for} reflection. 
    \revision{For example: ``I fixed a bug before submission''; or, ``We ended up having major merge conflicts''.}

    \item \textbf{Understanding.} The writer shows understanding of a goal, concept or personal experience, by interpreting it subjectively. This involves evaluating its significance, considering how it aligns with their desires or assumptions, and exploring what it means for them. The focus is on personal insights and high level evaluations, such as whether an experience was positive or negative, and why it happened, without delving into specific, concrete details. 
    \revision{For example: ``Teamwork was great, especially teammates did not hesitate to help each other''; or, ``Didn't code fast enough because I take so long doing research''.}

    \item \textbf{Feelings.} The writer identifies their feelings and emotions. This may include the students reported self-satisfaction with their behaviour (or lack thereof), in addition to experienced motivation or frustration. Students may discuss feelings like pride or happiness about their, or their team's, successes. Similarly, students may express frustration, annoyance, or disappointment arising from their, or the team's, actions. Such feelings may also arise from the emotional component of empathy (i.e., when emotions may be `contagious' between individuals \citep{waalPuttingAltruismBack2008}), but this must still be categorised by the individual describing their \textit{own} feelings or emotions. The other major component of empathy, cognitive empathy---where individuals can understand others' perspectives or mental states without experiencing their emotions for themselves \citep{rogersWhoCaresRevisiting2007}---corresponds to the `perspective' indicator below.
    \revision{For example: ``I feel a lot more motivated''; or, ``I feel like I have let down my team''.}

    \item \textbf{Reasoning.} The writer makes an in-depth analysis, leading to a deeper understanding of the experience by explaining in terms of concrete causes, effects, or consequences. This is distinct from the \textit{Understanding} indicator where the writer subjectively considers the experience from their own perspective, as this indicator emerges through a critical and objective analysis of concrete actions, events, or results. Similarly, this is distinct from the \textit{Hindsight} and \textit{Future Intention} indicators, in that this indicator extends only to analysing concrete causes for why something happened, and does not include the drawing of any conclusions from the experience. Such conclusions that take the form of considering how the situation could have been better approached would fall under \textit{Hindsight}, while specific actions the writer intends to take in the future correspond to \textit{Future Intention}.
    \revision{For example: ``I also took a dive into learning how thymeleaf fragments work which made for a lot less boilerplate code''; or, ``If I plan out my time better in the future I believe I'd be able to achieve more when it comes to tasks''.}

    \item \textbf{Perspective.} The writer acknowledges, considers, or discusses alternative perspectives. These perspectives may belong to their peers, team members, or instructors, and may have arisen through observation, discussion, or feedback. The perspectives can relate to a variety of topics, such as problems, experiences, solutions, emotions or challenges faced by others. For example, the writer might reflect on how a teammate struggled with a task, how a proposed solution was received by others, or how their actions impacted team morale. This indicator may arise as the result of the cognitive component of empathy \citep{rogersWhoCaresRevisiting2007}, where the writer understands another's perspective. However, any emotions that the \textit{writer} experiences themselves as a result of this empathy, would correspond to the `feelings' indicator. The perspectives must originate from someone other than the writer (e.g., a peer's opinion, a team member's difficulty, or an instructor's feedback), but they do not need to be in disagreement with the writer's own views. This indicator captures the writer's ability to step outside their own viewpoint and consider how others perceive or are affected by a situation, decision, or outcome.
    \revision{For example: ``After talking to other members in my team who agreed with my solution I see my mistake''; or, ``She had a hard time understanding as she hadn't yet worked with this part of the code base''.}

    \item \textbf{New Learning.} The writer describes what they have learned from the experience, in terms of either personal or professional skills. This may involve a better understanding of theory by putting it into practice, acquisition of new skills or abilities, or changes in perspective brought about by analysis of their experiences.
    \revision{For example: ``This gave me more knowledge about the database''; or, ``I've learned to communicate with my team in an effective way''.}

    \item \textbf{Hindsight.} The writer considers a past experience, and describes how---based on their new understanding as a result of reflection on the experience---they could, should, or would have approached the situation differently. This is commonly expressed through phrases such as ``should(n't) have'', or ``would(n't) have''.
    \revision{For example: ``I feel I should have helped more with the integration/unit tests''; or, ``I could've been more vocal in saying my opinions more right away instead of waiting for a moment where everyone is quiet''.}
    
    \item \textbf{Future Intention.} The writer indicates that they intend to do something and plans for the future. Here, ``something'' can be any future action, for example: trying a new process or changing their behaviour; talking to a peer; or, working on certain tasks. 
    \revision{For example: ``In the future, before sprint planning, I will spend extra time reading over the product backlog''; or, ``I should manage my time better, and when I complete a task, I should pick up another or check if the team needs any help''.}

\end{enumerate}

\label{sec:method:framework-design:examples}

Indicators are not mutually exclusive, and reflection texts may exhibit any combination of indicators. Consider the following example snippet from a reflection:

\begin{quote}
    \textit{I am proud of the work I did on the project. For the button to add images, I used CSS for the first time to correctly position and make the button transparent. }
\end{quote}

Here we can see two indicators of reflection present. In the first sentence ``I am proud of...'' the student expresses pride in their work, meaning that indicator 3 -- `Feelings' is present. In the second sentence, the student simply reports something that they did, meaning that indicator 1 -- `Description' is present. In this example the indicators apply to one sentence each, but indicators may also apply multiple times within a single sentence. Consider the following example:

\begin{quote}
    \textit{This week I learned how to mock dependencies in unit tests, which is something I really should have done in the tests I wrote last week, but now that I know how to I'll include it when testing the feature I am currently working on.}
\end{quote}

In this example we can see that the student begins by referencing some new learning that has taken place, meaning that indicator 6 -- `New Learning' is present. The student then includes both indicators 7 -- `Hindsight', and 8 -- `Future intention', when talking about something they should have been doing previously, and something they intended to do, respectively.

\section{Development and Evaluation of Classifier}
\label{sec:classifier-development}

\subsection{Annotation of Training Data for Reflection Classifier}
\label{sec:method:annotating-training-set}

To begin training an automated classifier for reflection indicators, we manually annotated a set of 1,518 \revision{responses} (totalling 6,704 sentences). \revision{Each of the 1,518 annotated texts corresponds to a single response to one of the three reflection prompts described in Section~\ref{sec:background:reflection-journaling-context}, rather than a complete weekly reflection (which contains three such responses).} \revision{The data collection context is described in Section~\ref{sec:method:data-collection-context}.} 
\revision{Only blank reflections were filtered out; no further filtering was performed on the dataset to ensure that our annotation captured the reality of reflections produced by students in a real course. Summary statistics for the annotated dataset---including word counts, indicator frequencies, and the distribution of indicators per response---are presented in Table~\ref{tab:dataset-statistics}.}

\begin{table}[htbp]
\centering
\caption{\revision{Summary statistics for the annotated dataset. Word counts and indicators per response are reported as mean (SD). Indicator frequencies show the number (and percentage) of responses in which each indicator was identified. The distribution of indicators per response shows how many responses contained 0, 1, 2, etc.\ indicators.}}
\label{tab:dataset-statistics}
\small
\begin{tabular}{lrrrr}
\hline
 & \textbf{Q1} & \textbf{Q2} & \textbf{Q3} & \textbf{Overall} \\
 & \textit{(did well)} & \textit{(did not do well)} & \textit{(do differently)} & \\
\hline
\multicolumn{5}{@{}l}{\textbf{Response characteristics}} \\
Words per response & 88.3 (73.7) & 81.7 (75.2) & 74.1 (78.2) & 82.3 (75.6) \\
Indicators per response & 2.94 (1.50) & 2.70 (1.57) & 2.19 (1.30) & 2.66 (1.51) \\
\hline
\multicolumn{5}{@{}l}{\textbf{Indicator frequencies (\% of responses)}} \\
Description & 68.6\% & 42.8\% & 5.1\% & 43.1\% \\
Understanding & 81.6\% & 81.0\% & 63.2\% & 76.8\% \\
Feelings & 52.0\% & 20.9\% & 7.5\% & 29.1\% \\
Reasoning & 26.8\% & 41.5\% & 33.6\% & 34.1\% \\
Perspective & 23.1\% & 21.2\% & 14.9\% & 20.4\% \\
New Learning & 39.0\% & 5.4\% & 2.7\% & 17.2\% \\
Hindsight & 0.7\% & 36.1\% & 3.5\% & 14.9\% \\
Future Intention & 2.3\% & 20.7\% & 88.5\% & 30.6\% \\
\hline
\multicolumn{5}{@{}l}{\textbf{Distribution of indicators per response}} \\
0 indicators & 6.9\% & 8.5\% & 9.6\% & 8.2\% \\
1 indicator & 11.2\% & 13.1\% & 20.3\% & 14.2\% \\
2 indicators & 19.5\% & 25.7\% & 30.9\% & 24.7\% \\
3 indicators & 25.0\% & 24.3\% & 26.4\% & 25.1\% \\
4 indicators & 20.7\% & 15.7\% & 8.3\% & 15.7\% \\
5 indicators & 14.2\% & 8.1\% & 2.9\% & 9.1\% \\
6 indicators & 2.5\% & 3.1\% & 1.3\% & 2.4\% \\
7 indicators & 0.0\% & 1.2\% & 0.3\% & 0.5\% \\
8 indicators & 0.0\% & 0.2\% & 0.0\% & 0.1\% \\
\hline
\end{tabular}
\end{table} 
Following guidelines for qualitative content analysis \citep{campbellCodingIndepthSemistructured2013}, the first author began by creating annotation guidelines for each of the eight indicators of reflection, to be applied to each of the three answers (full text) in a reflection, including: the criteria annotators use to determine whether each indicator was present in a piece of reflection text; exemplar statements of each indicator taken from real reflections;  common terms or words which typically do correspond with the presence of an indicator; and confounder terms or words which superficially may appear to correspond to an indicator, but actually do not. These annotation guidelines are included as supplementary material \citep{minishAssessmentSelfreflectionsSoftware2025}.

The initial set of guidelines produced by the first author was reviewed by
\revision{the second and third authors, and over several iterations the definitions, exemplars, and common / confounder phrases were refined to remove subjectivity and ambiguity between indicators where possible. This iterative refinement followed standard procedures for qualitative coding, as recommended by~\cite{mcdonaldReliabilityInterraterReliability2019}.}
Inter-rater agreement was then calculated by having three annotators (the first author, and two teaching staff of the project course the reflections were performed in) annotate 100 reflections texts using the criteria, and calculating the Krippendorff's alpha scores \citep{krippendorffContentAnalysisIntroduction2019} separately for each indicator, as this is a multi-label classification setting where indicators are completely independent.

After the first round of annotation, all disagreements were gathered and discussed by the three annotators. For each disagreement, a final decision was agreed upon, and the annotation guidelines for indicators with lower agreement scores were further refined to remove ambiguities and improve consistency. A second round of 100 annotations was then conducted to re-calculate agreement after these refinements. 

Following this process, annotators worked independently on different datasets to annotate a total of 1,518 \revision{responses} for training and evaluating the classifier. As a reliability check to verify annotation quality remained acceptable while annotators worked independently, one of every five such items was re-annotated by the first author and Krippendorff's alpha calculated.

\revision{Inter-rater agreement scores for the reflection annotation task are presented in Table~\ref{tab:inter-rater-agreement}. A Krippendorff's alpha greater than 0.8 typically indicates reliable rating, while 0.67--0.79 is often treated as a lower bound for tentative conclusions \citep{krippendorffContentAnalysisIntroduction2019}. Values below 0.67 are indicative of poor agreement, suggesting that the coding scheme is flawed, or is not being applied consistently by raters, while a value of 0 indicates no agreement better than chance \citep{krippendorffContentAnalysisIntroduction2019}.}

\revision{After the second round of agreement calculation, all Krippendorff's alpha scores were above the lower bound, and many were in the reliable range of 0.8+. Between human annotation rounds, inter-rater agreement improved for all indicators except future intention, which decreased from 0.96 to 0.89 (which still indicates high agreement). This decrease likely reflects natural variance in the annotation process where agreement could go up or down slightly when human annotators are involved \citep{krippendorffMeasuringReliabilityQualitative2004}. Still, the inter-rater agreement for future intention remains adequate, with the highest agreement score among all indicators. The agreement scores calculated from the reliability checks on independently labelled similarly increased, possibly as a result of annotators gaining more practice in applying the framework.}

\begin{table*}[htbp]
\centering
\caption{Inter-rater agreement for indicators by Krippendorff's alpha scores. Version 1 refers to the initial annotation guidelines after iterative refinement by the authors, while version 2 of the annotation guidelines includes changes made when reconciling disagreements after the first round of annotation using version 1. Also included are the agreement scores from the reliability check of independently labelled data, as described in Section~\ref{sec:method:annotating-training-set}.}
\label{tab:inter-rater-agreement}
\begin{tabular}{lrrr}
\hline
\textbf{Indicator} & \textbf{Version 1} & \textbf{Version 2} & \textbf{\makecell{Version 2\\(independent)}}  \\
\hline
Description & 0.69 & 0.87 & 0.88 \\
Understanding & 0.74 & 0.84 & 0.88 \\
Feelings & 0.62 & 0.82 & 0.93 \\
Reasoning & 0.64 & 0.72 & 0.80 \\
Perspective & 0.33 & 0.68 & 0.76 \\
New Learning & 0.69 & 0.79 & 0.87 \\
Hindsight & 0.71 & 0.85 & 0.90 \\
Future Intention & 0.96 & 0.89 & 0.88 \\
\hline
\end{tabular}
\end{table*} 
\subsection{\revisiontwo{Data Collection Context}}
\label{sec:method:data-collection-context}

\revision{The reflections were produced by 91 students at the University of Canterbury, New Zealand, taking a third-year software engineering project course in 2024. In the course, students worked in teams of six to eight on a year-long project following Scrum practices, with the project spanning six sprints of 2--4 weeks each over two semesters. Students were required to attend formal sessions in person on campus, though they were free to perform their development work either on campus or from home.}

\revision{Reflections were submitted weekly through a dedicated ``Weekly Self-Reflections'' interface in the ScrumBoard project management tool~\citep{minishScrumBoardProjectManagement2024} used by the course. The timing of submission within a given week was at students' discretion, though the typical pattern was to complete reflections on the Friday or over the weekend. Students received basic written guidance on how to write reflections through the course's student guide, but no further training (such as workshops or worked examples) was provided.}

\revision{Reflections were graded summatively at the end of each sprint, where teaching staff assigned a single traffic light grade (Red, Yellow, or Green) for the overall quality of a student's reflections within that sprint. Lateness or missing reflections could negatively affect this grade. As the unit of analysis for our annotation was the individual response to a single prompt rather than the sprint-level aggregate, these grades were not used in the present study.}

\revision{Across the 91 students, the number of annotated responses per student ranged from 11 to 30, with a mean of 16.7 (SD = 3.8, median = 16). Some variation is expected, as students could submit reflections of varying lengths and did not always answer all three prompts in every weekly reflection.}

\subsection{Reflection Classifier}
\label{sec:method:classifier}

For automated classification of reflections we explored two types of pre-trained language models: \emph{encoder-only} models such as BERT and its variants; and, \emph{decoder-only} large language models (LLMs) such as GPT or Qwen.

Encoder-only models like BERT have demonstrated promising performance on text classification tasks when fine-tuned with task-specific data \citep{sunHowFineTuneBERT2020}, with studies showing the fine-tuning allows such models to achieve state-of-the-art results across various NLP benchmarks \citep{sunHowFineTuneBERT2020}. 
In software engineering contexts, encoder-only models have been successfully applied to classify software requirements \citep{heyNoRBERTTransferLearning2020,kaurMNoRBERTMultilabelClassification2023}, identify bug report priorities \citep{izadiPredictingObjectivePriority2022}, and detect code smells \citep{alazbaCoRTTransformerbasedCode2024}. However, those models have not been used to classify reflective content in a software engineering context---see Section~\ref{sec:background:automated-approaches-for-assessing-reflective-writing} for their uses in reflection classification in other contexts.

Meanwhile, decoder-only models like GPT have shown excellent performance on few-shot learning tasks, scoring highly on many NLP benchmarks without gradient updates or fine-tuning \citep{brownLanguageModelsAre2020}. In software engineering, LLMs have been used for classification of software requirements \citep{binkhonainArePromptsAll2025}, software vulnerabilities \citep{luGRACEEmpoweringLLMbased2024}, and issues in open source projects \citep{aracenaApplyingLargeLanguage2025}. While LLMs remain largely unexplored as a means of classifying or assessing reflective texts, recent work in a computer science education context has suggested that LLM-guided reflections may be comparable to conventional reflections methods such as reflection questionnaires \citep{kumarSupportingSelfReflectionScale2024}.

Recent works comparing the two model architectures suggest that while fine-tuned encoder-only models generally outperform the prompting of decoder-only models when sufficient labelled data is available, generative decoder models like GPT can still match or approach a fine-tuned encoder's performance, particularly in low-data scenarios \citep{wangSelectingBERTGPT2024}.

To experiment with different models, we used the following:
\begin{itemize}
\item The encoder-only models were first downloaded from their public Huggingface\footnote{\url{https://huggingface.co/}}  repositories, and then fine-tuned on our annotated dataset with PyTorch\footnote{\url{https://pytorch.org/}}. 
\item The decoder-only models were also downloaded from Huggingface, and hosted locally via Ollama\footnote{\url{https://ollama.com/}}. 
\end{itemize}
This was all performed on the same machine, running Linux Mint 22, with 32GB of RAM, an Intel i7-14700 CPU, and Nvidia RTX 4070 12GB GPU. More information, including code artefacts used for training, hyper-parameter tuning, and automating the decoder-only models' classification, are included as supplementary material \citep{minishAssessmentSelfreflectionsSoftware2025}.

\subsubsection{Encoder-only Models}
\label{sec:method:classifier:encoder-only}

For encoder-only models, we explore BERT and its variants (e.g., RoBERTa, ALBERT, DeBERTa). These models are pre-trained on massive corpora and achieve strong performance on text classification tasks after fine-tuning \citep{devlinBERTPretrainingDeep2019,liuRoBERTaRobustlyOptimized2019,heDeBERTaDecodingenhancedBERT2021}. 

Our data set (see Section~\ref{sec:method:annotating-training-set}) consists of 1,518 annotated \revision{responses} (totalling 6,704 sentences), comparable to related works: 5,080 sentences across 77 essays \citep{ullmannAutomatedAnalysisReflection2019}; 7,128 sentences from 300 journals \citep{nehybaApplicationsDeepLanguage2023}; 1,113 sentences from 174 students \citep{alrashidiEvaluatingAutomatedAnalysis2023}.

Rather than training a new model from scratch---which would require learning both general language understanding and our specific classification task from our limited dataset---we leverage models already trained on general natural language and fine-tune them to focus specifically on our indicators. This approach has been used successfully in related works for classifying similar indicators of reflective writing in similarly sized training datasets  \citep{cavalcantiExplainablePredictionFeedback2023,nehybaApplicationsDeepLanguage2023}.

Since each reflection text can exhibit multiple reflection indicators simultaneously (see Section~\ref{sec:reflection-assessment-framework:indicators-of-reflection}), we frame this as a multi-label classification problem, and used a Binary Cross Entropy (BCE) loss function \citep{tsoumakasMiningMultilabelData2010}.
This treats each indicator as an independent binary prediction---consistent with how our annotators assessed each indicator separately during labelling (see Section~\ref{sec:method:annotating-training-set}).

To handle class imbalance of indicators in our dataset, we applied inverse frequency weighting to our loss function. Specifically, each class was weighted inversely proportional to its frequency in the training data, such that rarer indicators received proportionally higher weights during training. For example, if some indicator A appeared in 100 training samples, while indicator B appears in only 50 samples, indicator B would be weighted twice as strongly in the loss calculation. This ensures the model does not simply learn to predict only the most common indicators while ignoring rarer ones \citep{kingLogisticRegressionRare2001}.

To evaluate the classifier's performance, we employed 10-fold cross-validation \citep{kohaviStudyCrossvalidationBootstrap1995a}. The complete dataset of 1,518 annotated \revision{responses} was randomly divided into 10 equal folds. For each fold, we trained the model on 9 folds and evaluated on the remaining fold, repeating this process 10 times. This ensured that every reflection text was used for both training and testing, reducing the impact of any particular train-test split.

We compared the performance of five encoder-only models (BERT, DistilBERT, ALBERT, DeBERTa, and RoBERTa) using identical training configurations and cross-validation folds to ensure fair comparison. 
\revision{For all five models we used the base-sized variants available on Huggingface (\textbf{bert-base-uncased}, \textbf{roberta-base}, \textbf{distilbert-base-uncased}, \textbf{albert-base-v2}, and \textbf{microsoft/deberta-base}), to keep the comparison consistent across architectures and within the computational budget available for fine-tuning and hyperparameter tuning.} 
These five specific models were chosen for their widespread usage in text classification settings, beginning with the foundational model BERT \citep{devlinBERTPretrainingDeep2019}, and including its common variants with optimisations to: training approach \citep{liuRoBERTaRobustlyOptimized2019}; computational efficiency \citep{sanhDistilBERTDistilledVersion2020,lanALBERTLiteBERT2020}; and, architecture improvements \citep{heDeBERTaDecodingenhancedBERT2021}.

Based on the results of this comparison (detailed in Section~\ref{sec:results:model-comparison}), we selected the best-performing model and conducted hyperparameter tuning using the Optuna framework \citep{akibaOptunaNextgenerationHyperparameter2019} to further optimise classification performance.

During hyperparameter tuning we again used 10-fold cross-validation for evaluation during the tuning process. We optimised five key parameters: learning rate (log-uniform distribution between $1 \times 10^{-6}$ and $1 \times 10^{-4}$), training epochs (2-6), batch size (4, 8, 12, 16, 24, 32, 36, 48, or 64), positive class weight scaling (0.0-1.0), and classification threshold (0.3-0.7). The positive class weight scaling parameter controls the degree to which the inverse frequency weights (described above with regards to handling class imbalance) are applied: a value of 0.0 applies no class weighting (treating all classes equally), while 1.0 applies the full inverse frequency weighting, with intermediate values allowing for partial weighting adjustments.

These parameter ranges encompass standard configurations used by the original BERT authors \citep{devlinBERTPretrainingDeep2019}, and extend them with both higher and lower values to allow the tuning framework \citep{akibaOptunaNextgenerationHyperparameter2019} to empirically find the highest performing configurations for our context. The best hyperparameters from 1,000 trials were selected based on cross-validation performance, these results are reported in Section~\ref{sec:results:hyperparameter-tuning}.

\subsubsection{Decoder-only Models}
\label{sec:method:classifier:decoder-only}

In addition to encoder-only models, we explored decoder-only LLMs for reflection classification. Given the sensitive nature of student self-reflections and associated privacy and ethical considerations, we ran these models locally using Ollama \citep{ollamateamOllamaGetRunning2024} and publicly available models, rather than relying on third-party cloud services. 

We evaluated two open-source decoder-only models: OpenAI's gpt-oss 20B model \citep{openaiGptoss120bGptoss20bModel2025}; and, Alibaba's Qwen3 14B model \citep{yangQwen3TechnicalReport2025}. These models were selected for their strong performance on natural language understanding benchmarks, their manageable computational requirements for local deployment, and their open-source availability \citep{openaiOpenaiGptoss20bHugging2025,qwenteamQwenQwen314BHugging2025}.
For both models, we employed a zero-shot prompting approach to evaluate their capability to classify reflections using only the annotation guidelines, without requiring task-specific fine-tuning. \revision{We did not employ few-shot prompting (i.e., including labelled examples within the prompt), as our aim was to evaluate the decoder-only models' baseline capability to apply the annotation guidelines, mirroring the conditions under which human annotators initially approached the task. The potential effect of few-shot prompting on classification performance is discussed in Section~\ref{sec:discussion:limitations:internal-validity}.} The complete prompt and code artefacts used to automate the classification are available as supplementary material \citep{minishAssessmentSelfreflectionsSoftware2025}. This prompt contained all information given to human annotators (as described in Section~\ref{sec:method:annotating-training-set}), including example phrases and commonly related terms. The prompt explicitly instructed the model to treat each indicator independently, and to only consider was explicitly written in the reflection text without inferring anything. 

To evaluate the models, we prompted them to classify all 1,518 \revision{responses} by appending each text from the standardised prompt between the designated \texttt{[START\_CLASSIFICATION\_TEXT]} and \texttt{[END\_CLASSIFICATION\_TEXT]} delimiters. The models were instructed to return their classifications using a provided JSON object template with boolean values for each of the eight indicators, which we then parsed programmatically. 

Importantly, we enabled extended reasoning (or ``thinking'') modes for both models~\footnote{Via a flag in the model, see \url{https://huggingface.co/Qwen/Qwen3-14B#switching-between-thinking-and-non-thinking-mode}} to improve classification quality. Doing so causes the models to generate intermediate reasoning steps---effectively, thinking aloud for a time---before producing their final classification, which has been shown to improve performance on complex reasoning tasks \citep{weiChainofThoughtPromptingElicits2023}. 

We evaluated the decoder-only models using the same metrics as the encoder-only models to enable direct comparison: 
\begin{itemize}
    
\item General performance was assessed using Hamming loss and exact match accuracy.
\item Per-indicator performance was measured using accuracy, precision, recall, and F1 score. 
\end{itemize}
The results of the evaluations are presented in Section~\ref{sec:results:llm}. \section{Results}
\label{sec:results}

In this section we present: \revision{(1) a comparison of different transformer models fine-tuned on the annotated data; (2) evaluations of two final models after hyperparameter-tuning; and (3) the agreement of the final classifier with human annotators.}

\subsection{Encoder-only Model Comparison}
\label{sec:results:model-comparison}

\subsubsection{Comparison of Five Models}

\boxedtext{Results Summary:}{\revision{RoBERTa achieved the strongest overall performance among the five encoder-only models, with the lowest Hamming loss and consistently high F1-scores on the most challenging indicators (Perspective, Feelings, Reasoning). It was selected for hyperparameter tuning on this basis.}}

\revision{Table~\ref{tab:bert-models-f1-summary} summarises the F1-scores across all eight indicators for the five encoder-only models evaluated using 10-fold cross-validation. Full per-indicator results including precision, recall, and accuracy are provided in Appendix~\ref{appendix:full-model-comparison} (Table~\ref{tab:bert-models-comparison}).}

\begin{table}[htbp]
\centering
\caption{\revision{Encoder-only model F1-scores by indicator (mean ± std).
Full results including precision, recall, and accuracy are provided in
Appendix~\ref{appendix:full-model-comparison},
Table~\ref{tab:bert-models-comparison}.}}
\label{tab:bert-models-f1-summary}
\begin{tabular}{lrrrrr}
\hline
\textbf{Indicator} & \textbf{ALBERT} & \textbf{BERT} & \textbf{DistilBERT} & \textbf{DeBERTa} & \textbf{RoBERTa} \\
\hline
Description & \revision{0.724±0.060} & \revision{0.771±0.027} & \revision{0.777±0.027} & \revision{0.780±0.024} & \revision{\textbf{0.781±0.029}} \\
Understanding & \revision{0.863±0.145} & \revision{\textbf{0.916±0.022}} & \revision{0.916±0.018} & \revision{0.914±0.022} & \revision{0.914±0.020} \\
Feelings & \revision{0.663±0.070} & \revision{0.688±0.036} & \revision{0.695±0.059} & \revision{0.707±0.059} & \revision{\textbf{0.724±0.059}} \\
Reasoning & \revision{0.625±0.069} & \revision{0.671±0.031} & \revision{0.666±0.043} & \revision{\textbf{0.693±0.043}} & \revision{0.689±0.050} \\
Perspective & \revision{0.470±0.083} & \revision{0.492±0.062} & \revision{0.505±0.087} & \revision{0.524±0.080} & \revision{\textbf{0.537±0.133}} \\
New Learning & \revision{0.643±0.086} & \revision{0.701±0.076} & \revision{0.735±0.053} & \revision{0.764±0.054} & \revision{\textbf{0.812±0.063}} \\
Hindsight & \revision{0.791±0.138} & \revision{0.810±0.048} & \revision{0.824±0.053} & \revision{0.846±0.073} & \revision{\textbf{0.847±0.056}} \\
Future Intention & \revision{0.798±0.219} & \revision{0.889±0.040} & \revision{0.882±0.026} & \revision{0.897±0.027} & \revision{\textbf{0.909±0.029}} \\
\hline
\revision{Macro F1} & \revision{0.697±0.093} & \revision{0.742±0.016} & \revision{0.750±0.018} & \revision{0.766±0.018} & \revision{\textbf{0.777±0.029}} \\
\revision{Micro F1} & \revision{0.726±0.099} & \revision{0.776±0.016} & \revision{0.782±0.016} & \revision{0.791±0.014} & \revision{\textbf{0.802±0.015}} \\
Exact Match Acc. & \revision{0.242±0.070} & \revision{0.283±0.023} & \revision{0.302±0.043} & \revision{0.303±0.041} & \revision{\textbf{0.325±0.042}} \\
Hamming Loss & \revision{0.184±0.038} & \revision{0.157±0.015} & \revision{0.152±0.010} & \revision{0.148±0.012} & \revision{\textbf{0.136±0.010}} \\
\hline
\end{tabular}
\begin{center}
\footnotesize
\textbf{Note:} Best values for each metric are highlighted in bold.
\end{center}
\end{table}

\revision{All models showed strong performance on indicators with high human inter-rater agreement, with F1-scores generally exceeding 0.85 for Understanding and Future Intention across most models. However, performance varied substantially on the more challenging indicators--- Perspective, Feelings, and Reasoning---which also had lower inter-rater agreement between human raters (see Table~\ref{tab:inter-rater-agreement}). RoBERTa consistently achieved the highest or near-highest F1-scores on these challenging indicators. Given that all models performed satisfactorily on the high-agreement indicators, we prioritised the model that performed best where there was more room for improvement, and selected RoBERTa for hyperparameter tuning.}

\subsubsection{Final Model Performance}
\label{sec:results:hyperparameter-tuning}

\boxedtext{Results Summary:}{\revision{After hyperparameter tuning, the F1-optimised RoBERTa model improved exact match accuracy from 0.325 to 0.380. We provide two model variants: one optimised for overall accuracy (F1-optimised), suitable for general formative feedback where maximising feedback coverage is the priority; and one that favours recall, suitable for contexts where educators wish to minimise the risk of giving misleading feedback to students.}}

Following hyperparameter optimisation of the RoBERTa model,
\revision{Table~\ref{tab:final-performance-f1-summary} summarises the F1-scores for two model variants. Full per-indicator results including precision, recall, and accuracy are provided in Appendix~\ref{appendix:full-model-comparison} (Table~\ref{tab:final_performance}).} 
Here we show the performance of two models:

\begin{itemize}
    \item Firstly, we present a standard model with the highest mean F1-Score across all labels.
    \item Secondly, we present a model based on the highest mean score using the following heuristic which ``prefers'' recall whilst maintaining classification balance: $0.6 \times \text{recall} + 0.4 \times \text{F1}$. This configuration provides a ``safer'' alternative that educators might prefer if they are hesitant to rely on automated assessments of reflections when giving feedback to students, as it is more likely to give students the benefit of the doubt in ambiguous cases. We do this because feedback should only be given to students when it is relevant to their learning \citep{nicolFormativeAssessmentSelfregulated2006}, and poor quality or misleading feedback---that may arise from a misclassification by an automated assessment tool---can negatively affect students' self-esteem and performance \citep{blackAssessmentClassroomLearning1998}.

\end{itemize}

\begin{table}[ht!]
\centering
\caption{\revision{RoBERTa F1-scores by indicator after hyperparameter
tuning (mean ± std). Full results including precision, recall, and accuracy
are provided in Appendix~\ref{appendix:full-model-comparison},
Table~\ref{tab:final_performance}.}}
\label{tab:final-performance-f1-summary}
\begin{tabular}{lrrr}
\hline
\textbf{Indicator} & \textbf{\makecell{RoBERTa\\(before tuning)}} & \textbf{\makecell{RoBERTa\\(F1 Optimised)}} & \textbf{\makecell{RoBERTa\\(Recall Pref.)}} \\
\hline
Description & \revision{0.781±0.029} & \revision{\textbf{0.805±0.040}} & \revision{0.802±0.026} \\
Understanding & \revision{0.914±0.020} & \revision{0.930±0.014} & \revision{\textbf{0.931±0.013}} \\
Feelings & \revision{0.724±0.059} & \revision{\textbf{0.800±0.051}} & \revision{0.745±0.053} \\
Reasoning & \revision{0.689±0.050} & \revision{\textbf{0.732±0.043}} & \revision{0.691±0.050} \\
Perspective & \revision{0.537±0.133} & \revision{\textbf{0.595±0.048}} & \revision{0.541±0.041} \\
New Learning & \revision{0.812±0.063} & \revision{\textbf{0.844±0.034}} & \revision{0.749±0.049} \\
Hindsight & \revision{0.847±0.056} & \revision{\textbf{0.864±0.038}} & \revision{0.813±0.070} \\
Future Intention & \revision{\textbf{0.909±0.029}} & \revision{0.903±0.014} & \revision{0.886±0.037} \\
\hline
\revision{Macro F1} & \revision{0.777±0.029} & \revision{\textbf{0.809±0.013}} & \revision{0.770±0.025} \\
\revision{Micro F1} & \revision{0.802±0.015} & \revision{\textbf{0.831±0.011}} & \revision{0.793±0.022} \\
Exact Match Acc. & \revision{0.325±0.042} & \revision{\textbf{0.380±0.040}} & \revision{0.285±0.052} \\
Hamming Loss & \revision{0.136±0.010} & \revision{\textbf{0.119±0.011}} & \revision{0.162±0.026} \\
\hline
\end{tabular}
\begin{center}
\footnotesize
\textbf{Note:} Best values for each metric are highlighted in bold.
\end{center}
\end{table}  
Table~\ref{tab:confusion_matrices} shows confusion matrices for each indicator of the two final models. Here we can see that the model prioritising F1 score has the highest rate of true positive (top left cells) and true negative classifications (bottom left cells), while the recall weighted model sacrifices these metrics somewhat to minimise the rate of false negatives (top right cells). See Section~\ref{sec:discussion:readiness-for-real-use:two-model-approach} for an evaluation of what these values mean in practice.

\begin{table*}[htbp]
\centering
\caption{Confusion Matrices for RoBERTa Model Variants after Hyperparameter Tuning}
\label{tab:confusion_matrices}
\begin{tabular}{l|l|cc|cc}
\hline
\multirow{3}{*}{\textbf{Category}} & \multirow{3}{*}{\textbf{Predicted}} & \multicolumn{2}{c|}{\textbf{Recall Preference}} & \multicolumn{2}{c}{\textbf{F1 Optimised}} \\
\cline{3-6}
& & \multicolumn{2}{c|}{\textbf{Actual}} & \multicolumn{2}{c}{\textbf{Actual}} \\
& & \textbf{T} & \textbf{F} & \textbf{T} & \textbf{F} \\
\hline
\multirow{2}{*}{\textbf{Description}} & True & 613 & 275 & 570 & 213 \\
& False & 41 & 589 & 84 & 651 \\
\hline
\multirow{2}{*}{\textbf{Understanding}} & True & 1137 & 149 & 1136 & 137 \\
& False & 29 & 203 & 30 & 215 \\
\hline
\multirow{2}{*}{\textbf{Feelings}} & True & 393 & 248 & 376 & 147 \\
& False & 49 & 828 & 66 & 929 \\
\hline
\multirow{2}{*}{\textbf{Reasoning}} & True & 471 & 435 & 440 & 293 \\
& False & 46 & 566 & 77 & 708 \\
\hline
\multirow{2}{*}{\textbf{Perspective}} & True & 266 & 405 & 209 & 258 \\
& False & 43 & 804 & 100 & 951 \\
\hline
\multirow{2}{*}{\makecell[l]{\textbf{New} \\ \textbf{Learning}}} & True & 243 & 108 & 233 & 71 \\
& False & 18 & 1149 & 28 & 1186 \\
\hline
\multirow{2}{*}{\textbf{Hindsight}} & True & 207 & 59 & 198 & 42 \\
& False & 19 & 1233 & 28 & 1250 \\
\hline
\multirow{2}{*}{\makecell[l]{\textbf{Future} \\ \textbf{Intention}}} & True & 446 & 95 & 438 & 67 \\
& False & 19 & 958 & 27 & 986 \\
\hline
\end{tabular}
\begin{center}
\footnotesize
\textbf{Note:} For each category, T = True, F = False. Each 2×2 block shows True Positives, False Positives (left column) and False Negatives, True Negatives (right column).
\end{center}
\end{table*} 
\subsection{Decoder-only Model Comparison}
\label{sec:results:llm}

\boxedtext{Results Summary:}{\revision{Both decoder-only models substantially underperformed the fine-tuned RoBERTa model across all indicators, whilst requiring processing times two to three orders of magnitude longer (20--45 seconds
vs. under 0.1 seconds per classification). Qwen3:14b outperformed GPT-OSS:20b overall, but neither model is practical for real-time feedback applications.}}

Table~\ref{tab:llm-performance} presents the performance of two decoder-only LLMs evaluated on the complete dataset of 1,518 annotated \revision{responses} using zero-shot prompting with extended reasoning enabled (as described in Section~\ref{sec:method:classifier:decoder-only}).

\begin{table}[htbp]
\centering
\caption{Decoder-only Model Performance Comparison}
\label{tab:llm-performance}
\begin{tabular*}{\columnwidth}{@{\extracolsep{\fill}}lrr}
\hline
\textbf{Metric} & \textbf{GPT-OSS:20b} & \textbf{Qwen3:14b} \\
\hline
\multicolumn{3}{@{}l}{\textbf{Overall Performance}} \\
Exact Match Accuracy & 0.189 & \textbf{0.229} \\
Hamming Loss & 0.197 & \textbf{0.175} \\
\revision{Macro Precision} & \revision{\textbf{0.731}} & \revision{0.723} \\
\revision{Macro Recall} & \revision{0.785} & \revision{\textbf{0.795}} \\
\revision{Macro F1} & \revision{0.734} & \revision{\textbf{0.745}} \\
\revision{Micro Precision} & \revision{0.697} & \revision{\textbf{0.737}} \\
\revision{Micro Recall} & \revision{\textbf{0.790}} & \revision{0.785} \\
\revision{Micro F1} & \revision{0.741} & \revision{\textbf{0.760}} \\
\hline
\multicolumn{3}{@{}l}{\textbf{Description}} \\
Accuracy & 0.690 & \textbf{0.822} \\
Precision & 0.607 & \textbf{0.767} \\
Recall & \textbf{0.993} & 0.878 \\
F1-Score & 0.754 & \textbf{0.819} \\
\hline
\multicolumn{3}{@{}l}{\textbf{Understanding}} \\
Accuracy & 0.671 & \textbf{0.707} \\
Precision & 0.918 & \textbf{0.948} \\
Recall & 0.656 & \textbf{0.677} \\
F1-Score & 0.766 & \textbf{0.790} \\
\hline
\multicolumn{3}{@{}l}{\textbf{Feelings}} \\
Accuracy & \textbf{0.814} & 0.811 \\
Precision & \textbf{0.694} & 0.658 \\
Recall & 0.739 & \textbf{0.800} \\
F1-Score & 0.716 & \textbf{0.722} \\
\hline
\multicolumn{3}{@{}l}{\textbf{Reasoning}} \\
Accuracy & 0.650 & \textbf{0.696} \\
Precision & 0.508 & \textbf{0.550} \\
Recall & \textbf{0.971} & 0.868 \\
F1-Score & 0.667 & \textbf{0.673} \\
\hline
\multicolumn{3}{@{}l}{\textbf{Perspective}} \\
Accuracy & 0.811 & \textbf{0.822} \\
Precision & 0.615 & \textbf{0.633} \\
Recall & 0.384 & \textbf{0.424} \\
F1-Score & 0.473 & \textbf{0.508} \\
\hline
\multicolumn{3}{@{}l}{\textbf{New Learning}} \\
Accuracy & \textbf{0.920} & 0.913 \\
Precision & \textbf{0.747} & 0.707 \\
Recall & 0.878 & \textbf{0.889} \\
F1-Score & \textbf{0.807} & 0.788 \\
\hline
\multicolumn{3}{@{}l}{\textbf{Hindsight}} \\
Accuracy & \textbf{0.948} & 0.932 \\
Precision & \textbf{0.930} & 0.743 \\
Recall & 0.705 & \textbf{0.872} \\
F1-Score & \textbf{0.802} & \textbf{0.802} \\
\hline
\multicolumn{3}{@{}l}{\textbf{Future Intention}} \\
Accuracy & \textbf{0.924} & 0.899 \\
Precision & \textbf{0.827} & 0.780 \\
Recall & 0.953 & \textbf{0.957} \\
F1-Score & \textbf{0.886} & 0.860 \\
\hline
\end{tabular*}
\begin{center}
\footnotesize
\textbf{Note:} Best values for each metric are highlighted in bold.
\end{center}
\end{table} 
\subsubsection{GPT-OSS Performance}
\label{sec:results:llm:gpt-oss}

GPT-OSS:20b achieved an exact match accuracy of 0.189 and a Hamming loss of 0.197. The model demonstrated highly variable performance across indicators. 

Performance was strongest on indicators where human annotators also showed high agreement: Hindsight (F1: 0.802, accuracy: 0.948), Future Intention (F1: 0.886, accuracy: 0.924), and New Learning (F1: 0.807, accuracy: 0.920).

However, the model struggled significantly on some indicators. Perspective, also the indicator with the lowest inter-rater agreement ($\alpha = 0.68$), showed the weakest performance (F1: 0.473, accuracy: 0.811), with particularly low recall (0.384) indicating the model frequently failed to identify this indicator when present. Understanding also proved challenging, despite strong agreement between human raters ($\alpha = 0.84$), with the model achieving high precision (0.918) but relatively low recall (0.656), suggesting it was overly conservative in identifying this indicator.

\subsubsection{Qwen3 Performance}
\label{sec:results:llm:qwen3}

Qwen3:14b achieved an exact match accuracy of 0.229 and a Hamming loss of 0.175, representing improved overall performance compared to GPT-OSS:20b across both metrics.

The model demonstrated strong performance on several indicators with high human inter-rater agreement. Future Intention achieved the highest F1-score (0.860, accuracy: 0.899), followed by Description (F1: 0.819, accuracy: 0.822), Hindsight (F1: 0.802, accuracy: 0.932), and New Learning (F1: 0.788, accuracy: 0.913). 

However, challenges remained with certain indicators. Perspective, the indicator with the lowest inter-rater agreement ($\alpha = 0.68$), showed the weakest performance (F1: 0.508, accuracy: 0.822), though this represented an improvement over GPT-OSS's performance on the same indicator. The model achieved low recall (0.424), suggesting it remained conservative in identifying alternative perspectives. Understanding also struggled, despite strong agreement between human raters ($\alpha = 0.84$), with the model achieving high precision (0.948) but lower recall (0.677), indicating a tendency towards conservative classification similar to GPT-OSS but with somewhat improved recall performance.

\subsubsection{Comparison of Decoder-only Models}
\label{sec:results:llm:decoder-comparison}

Qwen3:14b performed better than GPT-OSS:20b across most metrics and indicators, achieving higher exact match accuracy (0.229 versus 0.189) and lower Hamming loss (0.175 versus 0.197). The performance differences were most pronounced on certain indicators: Qwen3:14b achieved notably higher F1-scores for Description (0.819 versus 0.754) and Understanding (0.790 versus 0.766), whilst both models struggled similarly with Perspective, achieving F1-scores of 0.508 and 0.473 respectively. For indicators with high human inter-rater agreement, both models performed well, with F1-scores near to or exceeding 0.8 for Hindsight (0.802 for both models), Future Intention (0.860 and 0.886), and New Learning (0.788 and 0.807).

\subsubsection{Comparison with Encoder-only Models}
\label{sec:results:llm:comparing-to-encoder-models}

Comparing the decoder-only models with the fine-tuned encoder-only models reveals clear differences in their performance for this task. The F1-optimised RoBERTa model achieved substantially better overall performance, with an exact match accuracy of \revision{0.380} compared to 0.229 for Qwen3:14b and 0.189 for GPT-OSS:20b. Similarly, RoBERTa's Hamming loss of \revision{0.119} was considerably lower than both Qwen3:14b (0.175) and GPT-OSS:20b (0.197), indicating more accurate multi-label predictions.

The performance gap between encoder-only and decoder-only models varied substantially across indicators. RoBERTa outperformed both decoder-only models on every single indicator when comparing F1-scores, which are derived from both precision and recall to provide a balanced measure of classification performance. For indicators with high human inter-rater agreement---such as Understanding, Hindsight, and Future Intention---both model types performed well, though RoBERTa consistently achieved higher F1-scores: Understanding (\revision{0.930} versus 0.790 and 0.766), Hindsight (\revision{0.864} versus 0.802 for both), and Future Intention (\revision{0.903} versus 0.860 and 0.886).

The differences were more pronounced for indicators where human annotators showed lower agreement. For Perspective, the indicator with the lowest human inter-rater agreement ($\alpha = 0.68$), RoBERTa achieved an F1-score of \revision{0.595} compared to 0.508 for Qwen3:14b and 0.473 for GPT-OSS:20b. Similarly, for Reasoning ($\alpha = 0.72$), RoBERTa's F1-score of \revision{0.732} exceeded both Qwen3:14b (0.673) and GPT-OSS:20b (0.667). These patterns suggest that fine-tuning on task-specific annotated data enabled the encoder-only models to better capture the nuanced distinctions that even human annotators find challenging, whilst zero-shot decoder-only models struggle more with these ambiguous cases despite their extended reasoning capabilities.

\subsubsection{Computational Efficiency for Practical Deployment}
\label{sec:results:llm:computational-efficiency}
 
\boxedtext{Results Summary:}{\revision{The fine-tuned RoBERTa model classifies a single reflection in under 0.1 seconds on consumer-grade hardware, making real-time feedback feasible. Decoder-only models required 20--45 seconds per classification, precluding real-time use.}}
 
\revision{For practical deployment in educational settings, computational efficiency is an important consideration alongside classification accuracy. Table~\ref{tab:decoder-model-timing-comparison} compares the processing times of the decoder-only models, both run on the same consumer-grade hardware as the encoder-only models (see Section~\ref{sec:method:classifier:encoder-only}).}

\begin{table}[htbp]
\centering
\caption{Decoder-only Model Timing Comparison (mean ± std). \revision{``Input processing'' refers to the time the model spends reading the annotation guidelines plus the reflection text to be classified. ``Response generation'' refers to the time the model spends generating its output (including extended reasoning).}}
\label{tab:decoder-model-timing-comparison}
\begin{tabular*}{\columnwidth}{@{\extracolsep{\fill}}lrr}
\hline
\textbf{Metric} & \textbf{GPT-OSS:20b} & \textbf{Qwen3:14b} \\
\hline
\multicolumn{3}{@{}l}{\textbf{Total Time per Sample}} \\
Time (seconds) & 44.6±28.1 & 20.7±8.0 \\
\hline
\multicolumn{3}{@{}l}{\textbf{Input Processing}} \\
Time (seconds) & 3.2±4.4 & 0.4±0.6 \\
\hline
\multicolumn{3}{@{}l}{\textbf{Response Generation}} \\
Time (seconds) & 36.1±19.8 & 20.3±8.0 \\
Tokens & 917±511 & 840±323 \\
\hline
\end{tabular*}
\end{table} 
\revision{Qwen3:14b averaged 20.7 seconds per classification, compared to GPT-OSS:20b's 44.6 seconds---a 54\% reduction in processing time. This difference was primarily driven by the response generation phase, where Qwen3:14b averaged 20.3 seconds compared to GPT-OSS:20b's 36.1 seconds. The high standard deviations in processing times (8.0 seconds for Qwen3:14b; 19.8 seconds for GPT-OSS:20b) reflect the varying complexity of reflection texts, which ranged from single sentences to multiple paragraphs.}

\revision{In contrast, the fine-tuned RoBERTa model classifies a single reflection text in 0.01--0.05 seconds---approximately three orders of magnitude faster than the decoder-only models. This difference has direct implications for deployment: encoder-only models can be used in real-time to provide feedback to students as they write or submit their reflections, whereas decoder-only models would introduce delays that may be impractical in an interactive educational setting.}

\subsection{Classifier Agreement with Human Annotations}
\label{sec:results:classifier-vs-human-annotators}

\boxedtext{Results Summary:}{\revision{When treated as a fourth annotator, the F1-optimised RoBERTa classifier maintained or exceeded human-level agreement on six of eight indicators. Description and Understanding showed lower agreement with human annotators than the other indicators.}}

To evaluate whether the F1-optimised RoBERTa classifier was able to approach human-level performance in applying the reflection assessment framework, we calculated Krippendorff's alpha treating the classifier as an additional rater alongside the three human annotators. This approach allows direct comparison between human-to-human agreement and classifier-to-human agreement on the same metric. Table~\ref{tab:inter-rater-agreement-with-classifier} presents these agreement scores across two datasets: the 100-sample set used in the final round of inter-rater agreement calculation (see Section~\ref{sec:method:annotating-training-set}); and, the complete dataset of 1,518 samples.

\revisiontwo{Throughout this comparison, the human-only reference baseline is the agreement achieved in the second round of joint annotation (the ``Version 2'' column of Table~\ref{tab:inter-rater-agreement}, reported here to three decimal places), as these values were calculated on the same 100-sample set, and with the same three annotators, as the classifier comparison. The higher agreement values observed during the reliability checks on independently annotated data (the ``Version 2 (independent)'' column of Table~\ref{tab:inter-rater-agreement})---e.g., $\alpha = 0.80$ for Reasoning and $\alpha = 0.76$ for Perspective---were calculated on a different sample and with a different pairing of raters, and are therefore not directly comparable with the agreement scores reported here.}

\begin{table*}[htbp]
\centering
\caption{Inter-rater agreement for indicators by Krippendorff's alpha scores. The second and third columns show agreement on the final 100-sample set used for determining inter-rater agreement (see Table~\ref{tab:inter-rater-agreement}), comparing human-only agreement with agreement when including the F1-optimised RoBERTa classifier as a fourth rater. The fourth column shows agreement across all 1,518 annotated samples. \revisiontwo{The human-only baseline corresponds to the ``Version 2'' column of Table~\ref{tab:inter-rater-agreement} (i.e., the second round of joint annotation, calculated on the same 100-sample set and with the same three annotators), reported here to three decimal places.}}
\label{tab:inter-rater-agreement-with-classifier}
\begin{tabular}{lrrr}
\hline
\textbf{Indicator} & \makecell{\textbf{Humans only} \\ \textbf{(per Table~\ref{tab:inter-rater-agreement})}} & \makecell{\textbf{Humans and} \\ \textbf{F1 RoBERTa} \\ \textbf{(n=100)}} & \makecell{\textbf{Humans and} \\ \textbf{F1 RoBERTa} \\ \textbf{(n=1,518)}} \\
\hline
\textbf{Description} & 0.872 & 0.773 & 0.787 \\
\textbf{Understanding} & 0.835 & 0.715 & 0.702 \\
\textbf{Feelings} & 0.822 & 0.828 & 0.922 \\
\textbf{Reasoning} & 0.719 & 0.720 & 0.843 \\
\textbf{Perspective} & 0.677 & 0.727 & 0.843 \\
\textbf{New Learning} & 0.787 & 0.820 & 0.916 \\
\textbf{Hindsight} & 0.849 & 0.886 & 0.955 \\
\textbf{Future Intention} & 0.889 & 0.913 & 0.959 \\
\hline
\end{tabular}
\end{table*} 
On the 100-sample set used in the final round of inter-rater agreement, when the classifier was included as a fourth rater, six of the eight indicators (Feelings, Reasoning, Perspective, New Learning, Hindsight, and Future Intention) maintained or exceeded the human-only agreement levels, with Perspective showing the most notable increase from 0.677 to 0.727. However, two indicators---Description and Understanding---showed decreases in agreement, dropping from 0.872 to 0.773 and 0.835 to 0.715 respectively. These decreases suggest that whilst the classifier performs well in absolute terms---achieving high accuracy, precision, and recall (see Table~\ref{tab:final_performance})---its judgements diverge more from human annotators on these indicators than on others.

When including the F1-optimised RoBERTa model as a fourth rater in the complete dataset of 1,518 samples, Krippendorff's alpha values ranged from $\alpha = 0.702$ to $\alpha = 0.959$. The highest agreement was observed for Future Intention ($\alpha = 0.959$), Hindsight ($\alpha = 0.955$), and Feelings ($\alpha = 0.922$), while Understanding showed the lowest agreement at $\alpha = 0.702$. Compared to the human-only agreement on the calibration set, most indicators maintained similar or higher alpha values on the full dataset. Specifically, Feelings (0.828 to 0.922), Reasoning (0.720 to 0.843), Perspective (0.727 to 0.843), New Learning (0.820 to 0.916), Hindsight (0.886 to 0.955), and Future Intention (0.913 to 0.959) all showed increases, while Description (0.787) and Understanding (0.702) remained lower than their respective human-only baselines (0.872 and 0.835). \section{Discussion}
\label{sec:discussion}

\revisiontwo{This section is organised around four threads. We first discuss the reliability of the reflective writing framework itself (Section~\ref{sec:discussion:annotation}) and situate our dataset relative to prior work (Section~\ref{sec:discussion:dataset}). We then interpret the performance of the classifiers and their agreement with human annotators (Section~\ref{sec:discussion:model-performance}). Building on these threads, Section~\ref{sec:discussion:readiness-for-real-use} consolidates the implications for real-world educational use---including the suitability of the framework, the choice between the two final models, and how indicator-level classification enables actionable feedback---and Section~\ref{sec:discussion:generative-ai} considers our contributions in the context of generative AI. We close with the limitations of this work (Section~\ref{sec:discussion:limitations}) and directions for future research that follow from the preceding threads (Section~\ref{sec:discussion:future-work}).}

\subsection{\revision{Reliability of the Reflective Writing Framework}}
\label{sec:discussion:annotation}

\boxedtext{Key Takeaway: }{\revision{The reflective writing framework can be applied reliably in practice: six of eight indicators achieved strong inter-rater agreement ($\alpha > 0.8$), whilst the remaining two exceeded the lower bound for tentative conclusions. The framework's modular design allows practitioners to include or exclude individual indicators based on their reliability requirements and pedagogical context.}}

The inter-rater agreement results presented in Table~\ref{tab:inter-rater-agreement} demonstrate that most indicators in our reflective writing framework achieved acceptable to strong levels of reliability after iterative refinement of the annotation guidelines. Six of the eight indicators exceeded the recommended threshold of $\alpha = 0.8$ for reliable coding \citep{krippendorffContentAnalysisIntroduction2019}, whilst the remaining two indicators (Reasoning and Perspective) achieved values above the lower bound of $\alpha = 0.67$ for tentative conclusions \citep{krippendorffContentAnalysisIntroduction2019}. This means that it is indeed possible to identify the indicators from our framework in reflection text. This confirms the practical applicability of the reflection framework and the indicators.

\subsubsection{\revision{Conceptual Validity of Indicators with Lower Agreement}}

The two indicators with lowest inter-rater agreement after the second round of joint annotation---Reasoning ($\alpha = 0.72$) and Perspective ($\alpha = 0.68$)---warranted careful consideration regarding their inclusion in the framework. 

For Reasoning, the primary challenge we encountered when defining annotation guidelines lay in distinguishing it conceptually from the Understanding indicator. While iterating on the guidelines, we clarified that Reasoning requires concrete cause-and-effect relationships---where the writer identifies that a specific, concrete action or event led to a particular, concrete outcome---whilst Understanding involves more subjective interpretation or high-level evaluation of experiences. For example, a student stating ``the merge conflicts occurred because we didn't communicate about which files we were working on'' demonstrates Reasoning through concrete causation, whilst ``I think communication is important for teamwork'' represents Understanding through subjective interpretation, often delivered by the writer as opinions without explicit links to concrete events. This distinction was refined through guideline development, and as annotators became more practised in applying the framework, we saw the inter-rater agreement rise to meet the `reliable' standard with a Krippendorff's $\alpha = 0.80$. Conceptually, the ability to distinguish between subjective (Understanding) and objective (Reasoning) analysis of one's experiences is key to separating between low levels of reflection, and critical reflection \citep{wongAssessingLevelStudent1995,boudPromotingReflectionLearning2013}. For this reason, and due to the increased agreement as annotation continued, we chose to retain Reasoning and Understanding as distinct indicators in the framework.

Perspective presented different challenges. As an infrequently occurring indicator in our dataset (present in only 20.4\% of texts; see Table~\ref{tab:dataset-statistics}), annotators had fewer opportunities to practice their judgements, which may have contributed to lower agreement. Through refining the criteria for the indicator to be less reliant on personal interpretation by the annotator, and clarifying that the `perspective' in question must pertain to someone \textit{other than} the author of the reflection, we were able to achieve a much higher---though still moderate---agreement between the first and second round of agreement checks. As annotation continued and annotators encountered more examples of this indicator, we saw from our reliability checks on the independently annotated data that agreement on the perspective indicator continued to improve, almost to the `reliable' standard. This, combined with the particular importance in a software engineering context that students reflect on the social and interpersonal aspects of their experiences, has led to us also retaining the Perspective indicator in our framework.

\subsubsection{Trade-offs between Comprehensiveness and Reliability}

We have opted to include indicators achieving only moderate reliability after the second round of inter-rater agreement. This was done for a few reasons:
\begin{itemize}
\item A framework restricted only to indicators with very high agreement ($\alpha > 0.8$) may be more reliable, but may also be less useful in real educational contexts due to the omission of important dimensions of reflection identified in established theories~\citep{deweyHowWeThink1910,schonReflectivePractitionerHow1983,kolbExperientialLearningExperience2014,mezirowTransformativeDimensionsAdult1991}. Both Reasoning and Perspective align with core aspects of reflective practice: critical analysis of experiences and consideration of alternative viewpoints are widely recognised as essential components of meaningful reflection~\citep{moonHandbookReflectiveExperiential2004}.

\item The observed agreement levels for Reasoning ($\alpha = 0.72$) and Perspective ($\alpha = 0.68$), whilst lower than for other indicators, remain above the threshold of $\alpha = 0.67$ typically considered acceptable for drawing tentative conclusions~\citep{krippendorffContentAnalysisIntroduction2019}. Moreover, the reliability checks on independently annotated data showed continued improvement, with agreement reaching $\alpha = 0.80$ for Reasoning and $\alpha = 0.76$ for Perspective (see Table~\ref{tab:inter-rater-agreement}). 
\revisiontwo{We note that these reliability-check values were calculated on a different sample (the independently annotated data) and with a different pairing of raters than the joint annotation rounds; when comparing the classifier against human annotators (Sections~\ref{sec:results:classifier-vs-human-annotators} and~\ref{sec:discussion:model-performance}), we therefore use the second-round joint annotation values as the reference baseline.}
This progression may show that as annotators gained experience applying the framework, their judgements became more consistent---with Reasoning ultimately achieving the threshold for reliable coding and Perspective approaching it. Given the inherently subjective nature of assessing reflective writing---where even expert annotators must make interpretive judgements about students' underlying thought processes---these agreement levels may be considered reasonable. The complexity of human reflection means that some dimensions will inevitably be more ambiguous to identify than others.

\item As the agreement levels suggest that only tentative conclusions should be drawn from these indicators, they should be treated with appropriate caution in practical applications. For instance, if the framework is used to inform feedback to students, instructors might prioritise the high-agreement indicators (Description, Understanding, Feelings, New Learning, Hindsight, and Future Intention) whilst treating assessments of Reasoning and Perspective more tentatively. Rather than assuming these classifications are definitively correct, they might be used to identify potential opportunities for feedback that warrant closer inspection. 

\end{itemize}

Importantly, our framework was designed such that all indicators are treated as completely independent. This means that practitioners who wish to exclude Reasoning or Perspective---whether due to concerns about reliability or because they are less relevant to their specific context---can do so without invalidating the assessment of the remaining indicators. 

\subsubsection{Comparison with Related Work Annotation Methods}
\label{sec:discussion:annotation:comparison-related-work}

\revisiontwo{Our approach to establishing inter-rater agreement differs from that of the related works whose frameworks we adapted (introduced in Section~\ref{sec:method:framework-design}), with implications for interpreting and comparing reliability metrics. We calculated Krippendorff's alpha across all annotated texts without pre-filtering.}

\revisiontwo{\cite{ullmannAutomatedAnalysisReflection2019} pre-filtered their dataset before calculating inter-rater agreement, retaining only sentences where a majority of crowdsourced annotators agreed on the classification, and reported Cohen's kappa values of 0.48--0.86 between group classifications aggregated through majority voting. \cite{nehybaApplicationsDeepLanguage2023} obtained initial Cohen's kappa values of only 0.10--0.51 across annotator pairs; in response, they likewise selected only sentences where the majority of annotators agreed, with their two most consistent annotators resolving the remaining 5.86\% of disagreements through consensus, ultimately reporting kappa values of 0.35--0.67 on the filtered and consensus-reached subset. \cite{alrashidiEvaluatingAutomatedAnalysis2023} employed three annotators and reported Cohen's kappa values of 0.46--0.75 across the reflection indicators, though individual values per indicator were not provided, and the paper does not explicitly describe whether any pre-filtering or iterative refinement occurred before calculating agreement. We also note that all three studies used Cohen's kappa---a metric designed for two raters---where metrics designed for multiple raters, such as Krippendorff's alpha or Fleiss' kappa, may have been more appropriate~\citep{krippendorffContentAnalysisIntroduction2019}.}

Agreement metrics calculated on pre-filtered data represent only the subset of cases where annotators already showed some consensus, excluding the most difficult or ambiguous cases where the framework may be hardest to apply consistently.
\revisiontwo{Such filtering may yield higher reliability estimates than what was actually observed during annotation, and means that reported indicator frequencies may not reflect the true occurrence rates in the original data, making it difficult to compare indicator frequency distributions across studies.}

Our more conservative approach---calculating Krippendorff's alpha on the complete unfiltered dataset---provides a more realistic assessment of how consistently the framework can be applied in practice. The strong agreement we achieved on most indicators ($\alpha > 0.8$ for six of eight indicators) demonstrates that our framework can be applied reliably even when including the most challenging cases. The moderate agreement on Reasoning ($\alpha = 0.72$) and Perspective ($\alpha = 0.68$), whilst lower than other indicators, still exceeds the threshold for tentative conclusions and represents genuine inter-rater reliability across the full spectrum of reflection texts, not just the easier cases.
 
\subsection{\revision{Dataset Characteristics in Context}}
\label{sec:discussion:dataset}

\boxedtext{Key Takeaway: }{\revision{Our dataset is drawn from real weekly reflections in a software engineering project course, differing from prior work in both context (project-based SE versus essays or general computer science) and unit of analysis (whole responses to structured questions versus individual sentences). This grounds our results in authentic student reflection produced under real course conditions.}}

Our dataset comprises 1,518 annotated \revision{responses} from third-year software engineering students at the University of Canterbury, New Zealand. These were weekly self-reflections in a group project course, structured around questions (see Section~\ref{sec:background:reflection-journaling-context}). The dataset contains 6,704 sentences with a mean length of 82 words per response (min=1, max=665, SD=76).

Table~\ref{tab:dataset-comparison} compares our dataset characteristics with previous work in reflective writing classification. Below we summarize key observations:

\begin{table*}[htbp]
\centering
\caption{Comparison of datasets used in reflection classification studies, including assessment frameworks, units of analysis, and text characteristics.}
\label{tab:dataset-comparison}

\textbf{This study}

\begin{tabular}{@{}p{2cm}p{\dimexpr\textwidth-3cm\relax}@{}}
\hline
\textit{Framework} & 8 boolean indicators: Description, Understanding, Feelings, Reasoning, Perspective, New Learning, Hindsight, Future Intention \\
\textit{Context} & Weekly reflections from third-year software engineering students in project course \\
\textit{Unit} & Whole response to 1 of 3 questions \\
\textit{Size} & 1,518 (6,704 sentences) \\
\textit{Length} & Mean 82 words (min=1, max=665, SD=76) \\
\hline
\end{tabular}

\vspace{1em}
\textbf{\cite{ullmannAutomatedAnalysisReflection2019}}

\begin{tabular}{@{}p{2cm}p{\dimexpr\textwidth-3cm\relax}@{}}
\hline
\textit{Framework} & Ordinal reflection score (1--3); 8 boolean indicators: Description of an experience, Feelings, Personal belief, Awareness of difficulties, Perspective, Outcome: Lessons learned, Outcome: Future intention \\
\textit{Context} & 77 reflective essays (67 from BAWE corpus, 10 from cited literature) \\
\textit{Unit} & Sentence \\
\textit{Size} & 5,080 sentences \\
\textit{Length} & 116,633 words total (mean: 23 words per sentence) \\
\hline
\end{tabular}

\vspace{1em}
\textbf{\cite{nehybaApplicationsDeepLanguage2023}}

\begin{tabular}{@{}p{2cm}p{\dimexpr\textwidth-3cm\relax}@{}}
\hline
\textit{Framework} & \textit{Same as \cite{ullmannAutomatedAnalysisReflection2019}} \\
\textit{Context} & Reflective journals from 300 Czech pre-service teachers \\
\textit{Unit} & Sentence \\
\textit{Size} & 7,128 sentences (of 33,859 total) \\
\textit{Length} & Mean 28.65 sentences per journal (SD=14.45); words per sentence not reported \\
\hline
\end{tabular}

\vspace{1em}
\textbf{\cite{alrashidiEvaluatingAutomatedAnalysis2023}}

\begin{tabular}{@{}p{2cm}p{\dimexpr\textwidth-3cm\relax}@{}}
\hline
\textit{Framework} & 7 boolean indicators: Description of an experience, Understanding, Feelings, Reasoning, Perspective, Future action, New learning \\
\textit{Context} & Reflective writing from 74 third- and fourth-year CS students from a UK university \\
\textit{Unit} & Sentence \\
\textit{Size} & 1,113 sentences \\
\textit{Length} & Not reported \\
\hline
\end{tabular}

\end{table*} 
\begin{itemize}
\item \textbf{Reflection context:} Our semi-structured three-question format (see Section~\ref{sec:background:reflection-journaling-context}) differs from the reflective essays used by \cite{ullmannAutomatedAnalysisReflection2019}, and the free-form journals used by \cite{nehybaApplicationsDeepLanguage2023}. These three questions directly influence the nature of indicators present in responses. For example, Question 3 solicits the Future Intention indicator, whilst Questions 1 and 2 prompt more retrospective content, such as the Description or Hindsight indicators.
The context of our reflections differs slightly from Alrashidi et al.'s third- and fourth-year computer science students in the UK \citep{alrashidiEvaluatingAutomatedAnalysis2023}, and substantially from Nehyba and \v{S}tef\'{a}nik's Czech pre-service teachers \citep{nehybaApplicationsDeepLanguage2023} and Ullmann's essay corpus \citep{ullmannAutomatedAnalysisReflection2019}. Software engineering students reflecting on project work may emphasise technical problem-solving and team dynamics differently than other populations. 

\item \textbf{Unit of analysis:} We chose to classify complete responses to reflection prompts, whilst \cite{ullmannAutomatedAnalysisReflection2019} classified individual sentences, with \cite{nehybaApplicationsDeepLanguage2023}, and \cite{alrashidiEvaluatingAutomatedAnalysis2023} following this same approach. 
Our reasoning for analysing whole responses, rather than individual sentences, is as follows. First, our reflection format is already split into distinct themes via the three-questions used (see Section~\ref{sec:background:reflection-journaling-context}). 
Second, treating sentences in isolation loses important context for identifying reflective elements--- \cite{nehybaApplicationsDeepLanguage2023} found that classification performance improved when sentences were analysed alongside their surrounding context. 
Finally, as human educators we consider the full context of a reflection---not isolated sentences---when giving feedback, so we chose to train our classifier in the same way.
\end{itemize}
 
\subsection{\revision{Model Performance and Human Agreement}}
\label{sec:discussion:model-performance}

\boxedtext{Key Takeaway: }{\revision{Classifier performance closely mirrors human inter-rater agreement: indicators that humans classify consistently are also classified well by the models, whilst indicators with lower human agreement remain challenging for automated approaches. The fine-tuned RoBERTa encoder-only model substantially outperforms decoder-only models in both accuracy and speed, making it suitable for real-time deployment.}}

\subsubsection{\revision{Classifier Performance Varies by Indicator Difficulty}}

The performance of our classifiers varied substantially across indicators in patterns that closely mirror human inter-rater agreement, suggesting that indicators which humans find difficult to classify consistently also present challenges for automated approaches.

Indicators with high human inter-rater agreement---including Description, Understanding, Feelings, New Learning, Hindsight, and Future Intention---achieved strong classification performance across all model types, with the F1-optimised RoBERTa model \revision{achieving F1-scores of 0.80 or higher for all six of these indicators} (see Table~\ref{tab:final_performance}). These results suggest that when human annotators can apply the framework consistently, machine learning models can learn to replicate those judgements.

Conversely, the two indicators with lowest human agreement---Reasoning and Perspective---proved most challenging for classifiers (see Table~\ref{tab:final_performance}). These patterns are not surprising: if expert human annotators equipped with detailed guidelines struggle to consistently identify them, we should not expect machine learning models to resolve these ambiguities definitively. The moderate performance on these indicators likely reflects genuine ambiguity in student writing, or the assessment framework itself, rather than classifier limitations.

\subsubsection{\revision{Encoder-only Models Enable Real-time Deployment}}

The stark differences in computational efficiency between encoder-only and decoder-only models have direct implications for practical deployment in educational settings.

Our fine-tuned RoBERTa encoder-only model classifies individual reflection texts in 0.01–0.05 seconds on standard, consumer-grade hardware---making real-time feedback feasible. A student's reflection could be analysed, and feedback given without wait. This aligns with formative assessment principles, where timely feedback is preferred \citep{nicolFormativeAssessmentSelfregulated2006}.

In contrast, the decoder-only models---whilst requiring no task-specific fine-tuning---averaged 20.7 seconds (Qwen3:14b) to 44.6 seconds (GPT-OSS:20b) per classification. These processing times, whilst acceptable for small-scale analysis or research purposes, preclude real-time feedback applications.

This performance gap suggests different use cases for the two approaches. Encoder-only models are appropriate for production deployment where real-time or near-real-time feedback is desired and where sufficient annotated data exists for fine-tuning. Decoder-only models may be more suitable for prototyping new frameworks, analysing small datasets where fine-tuning is impractical, or in contexts where zero-shot flexibility outweighs processing time.

The zero-shot capability of decoder-only models does offer advantages in adaptability. Changes to indicator definitions or addition of new indicators require only prompt modifications rather than retraining, potentially reducing the iteration time when refining assessment frameworks. However, our results suggest this flexibility comes at the cost of both classification accuracy and computational efficiency. \revisiontwo{Here we restrict ourselves to interpreting these efficiency results; the broader implications for deploying the classifier in educational practice---including how educators may choose between and configure the two final models---are consolidated in Section~\ref{sec:discussion:readiness-for-real-use}.}

\subsubsection{\revision{Classifier Approaches Human-level Agreement on Most Indicators}}

To evaluate whether our classifier approaches human-level performance, we calculated Krippendorff's alpha treating the F1-optimised RoBERTa model as an additional annotator (see Table~\ref{tab:inter-rater-agreement-with-classifier}). This allows direct comparison between classifier-to-human agreement and human-to-human agreement. \revisiontwo{As in Section~\ref{sec:results:classifier-vs-human-annotators}, the human-only baseline used throughout this comparison is the agreement achieved in the second round of joint annotation (the ``Version 2'' column of Table~\ref{tab:inter-rater-agreement}), since these values were calculated on the same 100-sample set and with the same three annotators; the higher values from the reliability checks on independently annotated data (see Section~\ref{sec:discussion:annotation}) are not used as the baseline, as they arise from a different sample and rater pairing.}

For several indicators, classifier-to-human agreement approached or exceeded human-to-human agreement levels (see Table~\ref{tab:inter-rater-agreement-with-classifier}). These results suggest that for high-agreement indicators, the classifier performs comparably to an expert human annotator.

The two indicators with lowest human-only agreement showed interesting patterns when the classifier was included. For Perspective, agreement improved notably: from $\alpha = 0.68$ (human-only) to $\alpha = 0.73$ when including the classifier on the final agreement set, and further to $\alpha = 0.84$ across the complete dataset. For Reasoning, human-only agreement remained relatively stable at $\alpha = 0.72$ on both the final agreement set and when including the classifier, then improved to $\alpha = 0.84$ across the complete dataset. Whilst these values exceed the threshold for tentative conclusions, the initially moderate agreement indicates these remain the most challenging indicators to classify consistently.

The pattern for Description and Understanding indicators is noteworthy. When including the classifier on the final agreement set, agreement decreased slightly for Description and Understanding (see Table~\ref{tab:inter-rater-agreement-with-classifier}). This suggests that whilst the classifier performs well in absolute terms---achieving high accuracy, precision, and recall (see Table~\ref{tab:final_performance})---its judgements diverge more from human annotators on these indicators than on others. The classifier may have learned patterns in the training data that, whilst consistent and predictive, do not fully align with how expert humans apply the framework for these particular dimensions.
 
\subsection{Readiness for Real Educational Applications}
\label{sec:discussion:readiness-for-real-use}

\boxedtext{Key Takeaway: }{\revision{The classifier enables a shift from broad, holistic feedback on reflections to structured, indicator-level feedback that identifies specific dimensions a student may have overlooked. Educators can determine which indicators they expect for a given context, and deliver feedback on missing indicators---for example, as nudges that guide students towards deeper reflection without prescribing what to write. Two model variants support different pedagogical priorities, and the framework's independence of indicators allows educators to tailor which dimensions are assessed and how feedback is delivered.}}

As discussed in Sections~\ref{sec:discussion:annotation:comparison-related-work} and~\ref{sec:results:classifier-vs-human-annotators}, our framework not only achieved strong inter-rater agreement without pre-filtering data, but classifier models demonstrated agreement with human annotators comparable to expert-level performance across most indicators.

\subsubsection{\revisiontwo{Framework Suitability for Real-world Use}}
\label{sec:discussion:annotation:suitability-for-real-world-use}

\revisiontwo{Building on the reliability results discussed in Section~\ref{sec:discussion:annotation}, the tiered reliability of indicators has direct consequences for how the framework should be used in practice.}
The high-agreement indicators (Description, Understanding, Feelings, New Learning, Hindsight, and Future Intention) can be classified with substantial confidence, enabling reliable automated feedback on these dimensions. The moderate-agreement indicators (Reasoning and Perspective) may require more cautious interpretation, with automated classifications highlighting potential opportunities for feedback, rather than definitive assessments.

This tiered approach to reliability is appropriate for the intended use case of providing formative feedback in educational settings. The goal is not to replace instructor assessment, but to enable scalable, timely feedback that can guide students' ongoing reflection throughout a course. In this context, a framework that captures important but difficult to assess dimensions of reflection---even with moderate reliability---may be more valuable than a narrower but more reliable framework that misses key aspects of reflective practice. For example, educators may integrate automated feedback to students as they are writing their reflections, but only show feedback relating to the higher reliability indicators at first. Similarly, by knowing which indicators are more and less reliable, educators can tailor the tone of feedback messages: e.g., ``You have not discussed X'' for high reliability indicators, but ``Our system \revisiontwo{finds} you may not have discussed X'' for low reliability indicators. 

\subsubsection{Two-Model Approach for Different Use Cases}
\label{sec:discussion:readiness-for-real-use:two-model-approach}

Given the difficulty of robustly identifying indicators in reflective writing, where even human experts do not show perfect agreement, we provide two classifier models for educators to choose from. 

The F1-optimised model maximises overall accuracy to most closely match classification by a human expert, while the recall-optimised model gives a more conservative approach, useful in cases where educators may strongly wish to avoid giving misleading feedback (i.e., suggesting a student discuss an indicator of reflection that they have already addressed, but was missed during classification). 

In a practical deployment for generating formative feedback---where a missing indicator would trigger a prompt to the student---the classification outcomes shown in Table~\ref{tab:confusion_matrices} have the following implications:

\begin{itemize}
    \item \textbf{True Positives} (Top left in cell): The student has addressed the indicator; no feedback is given. (Correct decision)
    \item \textbf{True Negatives} (Bottom right in cell): The student has not addressed the indicator; feedback is given. (Correct decision)
    \item \textbf{False Positives} (Bottom left in cell): The student has not addressed the indicator, but no feedback is given. (Missed opportunity)
    \item \textbf{False Negatives} (Top right in cell): The student has addressed the indicator, but receives feedback anyway. (Potentially confusing feedback)
\end{itemize}

The recall-preference model (Table~\ref{tab:confusion_matrices}) correctly identifies 6,330 opportunities for formative feedback (true negatives summed across all indicators), at the cost of 264 instances of potentially confusing feedback (false negatives).

The F1-optimised model identifies 6,876 feedback opportunities (true negatives summed across all indicators)---546 more than the recall-preference model---but produces 440 false negatives, 176 more instances of potentially confusing feedback. This represents a trade-off between feedback coverage and precision: the F1-optimised model provides more opportunities for feedback, but with an increased risk of potentially confusing students, whilst the recall-preference model is more conservative but may miss opportunities for feedback.

\subsubsection{Guidance for Software Engineering Training}

Our assessment framework and automated classifier are designed for formative assessment---supporting ongoing learning rather than determining grades. The classification identifies concepts students may have missed in their reflections, but instructors retain control over pedagogical decisions, including which indicators to include, how they \revisiontwo{are} prioritised, and how any guidance is presented to students, e.g., immediate feedback messages; periodic summary reports; or, flagging reflections for instructor review.

The approach offers practical advantages for formative assessment: fast classification on every-day computer hardware enables feedback in-action rather than retrospectively; automated classification allows integration with large cohorts without prohibitive instructor workload; and the indicators provide a straightforward framework that educators can build on to suit their specific pedagogical context.

This approach should not be used for summative assessment. It was not designed to support summative assessment, and we feel using automated classification to determine marks would be inappropriate given the inherent subjectivity in assessing reflection and the potential for students to game the system if classifications directly affect grades.

\subsubsection{\revision{Enabling Actionable Feedback Through Indicator-level Classification}}
\label{sec:discussion:actionable-feedback}

\revision{A central motivation of this work is addressing the limitation that feedback on reflective writing tends to be broad and non-specific (see Section~\ref{sec:introduction}). Our framework and classifier address this by enabling feedback at the level of individual reflection indicators, rather than at the level of overall reflection quality.}

\revision{In practice, this operates as follows. Educators first determine which indicators of reflection they consider appropriate for a given context---for instance, they may expect responses to Question 3 (``What will you do differently in future?'') to exhibit the Future Intention indicator, and potentially Reasoning to justify the intended change. The classifier then identifies which of the expected indicators are absent from a student's response. Feedback can be generated for each missing indicator, directing the student's attention to specific dimensions of the reflection they may have overlooked.}

\revision{One approach to delivering such feedback is through nudges, drawing on the principles of nudge theory and choice architecture~\citep{thalerNudgeImprovingDecisions2008}. In this approach, the learning environment is structured to guide students towards deeper reflection without restricting their options or prescribing what they should write~\citep{dimitrovaChoiceArchitectureNudges2022}. For example, a student who has described what they will do differently, but has not provided reasoning, might receive a prompt to consider \textit{why} they believe this change will be effective. Such nudges have been applied in computing education to support the quality of reflective writing~\citep{mohammadhassanDiscoveringDifferencesLearning2022,mohammadhassanInvestigatingEffectNudges2022}, and are well-suited to reflection in software engineering project-based learning, where reflection is an inherently personal and self-directed activity that may be undermined by overly prescriptive feedback. Alternatively, educators may prefer to use the classifications to generate periodic summary reports highlighting patterns across multiple reflections, or to flag specific reflections for manual review.}

\revision{This indicator-level approach represents a qualitative shift from the status quo. Rather than receiving feedback such as ``your reflection could be deeper''---which leaves students uncertain about what to improve---a student might instead receive a prompt to consider the perspectives of their team members, or to articulate concrete reasons for why an experience unfolded as it did. The framework thus translates the abstract goal of ``deeper reflection'' into concrete, identifiable dimensions that students can act upon.}

\subsection{\revisiontwo{Reflection and Automated Assessment in the Era of Generative AI}}
\label{sec:discussion:generative-ai}

\boxedtext{Key Takeaway: }{\revisiontwo{Structured reflection may become more, rather than less, important as generative AI tools reshape how students learn: reflection asks students to make personal meaning of their own experiences, and indicator-level feedback prompts further thinking rather than producing content on students' behalf. At the same time, AI-based educational methods---including ours---carry risks of misclassification and bias, and some institutions restrict their use; deployments should therefore remain formative, transparent, and under instructor control.}}

\subsubsection{\revisiontwo{Structured Reflection and Overreliance on Generative AI}}

\revisiontwo{Our dataset was collected in 2024, when generative AI tools were already widely available to students, and the role of such tools in software engineering education has continued to evolve since~\citep{pratherRobotsAreHere2023,kasneciChatGPTGoodOpportunities2023}. A recurring concern in this context is over-reliance: when students delegate cognitive work to generative AI tools, the engagement with material that underpins learning may be reduced, with a growing body of work associating over-reliance on AI dialogue systems with negative effects on skills such as critical thinking and self-regulation~\citep{zhaiEffectsOverrelianceAI2024,kasneciChatGPTGoodOpportunities2023}.}

\revisiontwo{We believe structured reflection may act as a partial counterbalance to this tendency. The reflection format studied here asks students to make personal meaning of their own recent experiences---what they did, how they felt, how others were affected, and what they intend to do next---which is an inherently metacognitive activity linked to self-regulated learning~\citep{zimmermanBecomingSelfRegulatedLearner2002}, and one whose value to the student lies in the thinking itself rather than in the written artefact produced. Moreover, the feedback approach described in Section~\ref{sec:discussion:actionable-feedback} prompts students to think further about dimensions they have overlooked, rather than generating reflective content on their behalf; this is consistent with recent findings that carefully scoped LLM-guided reflection can support, rather than supplant, students' own reflective processes~\citep{kumarSupportingSelfReflectionScale2024}. That said, students may nonetheless use generative AI tools to draft their reflections. Our classifier identifies indicators of reflection in text; it does not verify that the underlying reflective process took place, and distinguishing genuine reflection from AI-generated text remains an open problem that we highlight as a direction for future work (Section~\ref{sec:discussion:future-work}).}

\subsubsection{\revisiontwo{Risks and Biases of AI-based Educational Tools}}

\revisiontwo{The adoption of AI-based tools for course management and assessment also carries risks that warrant consideration, particularly given that some educational institutions restrict or prohibit their use~\citep{miaoGuidanceGenerativeAI2023,kasneciChatGPTGoodOpportunities2023}. Beyond the misclassification risks discussed in Section~\ref{sec:discussion:readiness-for-real-use:two-model-approach}, machine learning models trained on data from one population may perform systematically worse for particular subgroups of students---for example, those with different cultural backgrounds or levels of English proficiency---a form of algorithmic bias documented across educational applications of AI~\citep{bakerAlgorithmicBiasEducation2022}. As our classifier was trained on data from a single institution (see Section~\ref{sec:discussion:limitations:external-validity}), such biases cannot be ruled out and should be evaluated before deployment in new contexts.}

\revisiontwo{Several aspects of our approach are intended to mitigate these risks. The classifier runs locally on consumer-grade hardware, so student reflections need not be shared with third-party services (see Section~\ref{sec:method:classifier:decoder-only}); its per-indicator outputs are transparent rather than an opaque holistic judgement; the two model variants allow educators to manage the trade-off between feedback coverage and misleading feedback (Section~\ref{sec:discussion:readiness-for-real-use:two-model-approach}); and we position the tool for formative use only, with instructors retaining control over which indicators are assessed and how feedback is delivered. Nonetheless, any deployment should comply with institutional policies on AI-based tools, be transparent to students that feedback is automatically generated, and retain instructor oversight.}

\subsection{Limitations}
\label{sec:discussion:limitations}

Our study has some limitations to consider when interpreting the results and applying the framework in practice. These limitations are reported in line with the framework presented in \cite{wohlinExperimentationSoftwareEngineering2012}.

\subsubsection{Conclusion Validity}
\label{sec:discussion:limitations:moderate-agreement-indicators}

Two indicators---Reasoning ($\alpha = 0.72$) and Perspective ($\alpha = 0.68$)---showed only moderate inter-rater agreement amongst human annotators. Whilst these values exceed the threshold for tentative conclusions, and match or exceed the agreement reported in related works, classifications for these indicators should be interpreted more cautiously.

\subsubsection{Internal Validity}
\label{sec:discussion:limitations:internal-validity}

Our study is subject to several confounding factors. The prompts given to decoder-only models (GPT, Qwen) may have influenced their performance. Whilst we provided the same annotation guidelines to both decoder-only models and human annotators, different prompt structures or phrasings might yield different results. \revision{In particular, we used zero-shot prompting (providing only the annotation guidelines, without labelled examples), as this allowed us to evaluate baseline model capability without confounding the comparison with example selection effects. Few-shot prompting---where a small number of labelled examples are included in the prompt---has been shown to improve performance on similar classification tasks~\citep{brownLanguageModelsAre2020}, and may have produced stronger results for the decoder-only models. We did not explore this systematically here, but flag it as a consideration when interpreting our results.} Additionally, we used versions of these models (GPT-OSS:20B and Qwen3:14B) that could run on consumer-grade hardware, whereas larger variants (e.g., GPT-OSS:120B, Qwen3:235B) might achieve better classification performance, though would require significantly greater computing resources. 

During annotation guideline refinement, discussions between annotators may have established implicit criteria that were not fully captured in the final written guidelines. Whilst this tacit knowledge would be shared between human annotators, it remains unavailable to the decoder-only models that relied on the documented guidelines, potentially limiting their performance when compared to human annotators or the encoder-only models.

\subsubsection{External Validity}
\label{sec:discussion:limitations:external-validity}

Our classifier was trained and evaluated exclusively on reflections from students at a single software engineering project course at the University of Canterbury, New Zealand. This limits generalisability in several ways.

First, our three-question reflection format (see Section~\ref{sec:background:reflection-journaling-context}) may influence which indicators appear in student responses. Whilst the classifier was trained on responses to all three questions, its performance on fully free-form reflections or differently structured prompts has not been assessed.

Second, differences in student populations (e.g., cultural backgrounds, prior education, English language proficiency) or institutional practices (e.g., teaching and assessment methods) could affect classifier performance. The classifier's generalisability to different educational contexts---such as alternative reflection formats, different student populations, or non-project-based courses---remains to be evaluated.

Finally, our dataset size of 1,518 \revision{responses}, whilst comparable to related work (see Table~\ref{tab:dataset-comparison}), may limit the model's ability to transfer to other reflective writing styles. Validation on larger and more diverse datasets would strengthen confidence in the classifier's generalisability.

\subsubsection{Construct Validity}

Whilst our indicators are derived from established reflection theories, it remains to be seen whether they capture all important elements of reflection in software engineering education contexts. The framework may miss discipline-specific aspects of reflection that are particularly relevant to software development practice  such as reflection on technical decision-making and trade-offs, learning from code reviews or debugging experiences, or consideration of software quality attributes and their impact on design choices. 

We classify complete responses to individual reflection questions rather than individual sentences. This approach suits our structured reflection format, and is suited to providing holistic feedback as a human educator typically would, but may not be suitable for contexts where per-sentence analysis is desired.

\subsection{Future Work}
\label{sec:discussion:future-work}

Several directions for future research \revisiontwo{follow from the threads developed in this discussion: the deployment of the classifier and the design of feedback around it (following Sections~\ref{sec:discussion:readiness-for-real-use} and~\ref{sec:discussion:generative-ai}); refinement of the reflective writing framework itself (following Section~\ref{sec:discussion:annotation}); and validation and generalisation beyond our dataset (following Section~\ref{sec:discussion:dataset} and the limitations above). We present the directions below in this order.}

\subsubsection{Real-world Deployment and Evaluation}

The most immediate next step is deploying the classifier in an active software engineering project course\revisiontwo{, realising the deployment scenarios outlined in Section~\ref{sec:discussion:readiness-for-real-use}}. \revision{Such a deployment requires the design of a feedback delivery system that translates classifier outputs into pedagogically appropriate messages for students. This is a non-trivial undertaking, involving decisions about which indicators to assess for each reflection question, how feedback should be worded and timed, and what form it should take---for instance, whether as real-time nudges during writing, post-submission summaries, or flags for instructor review. These design decisions carry pedagogical implications that warrant careful consideration and evaluation in their own right.} This would enable evaluation of how students respond to automated feedback, whether feedback supports improvement in reflection quality over time, and provide a practical \revision{guide} for educators to follow in their own courses.

\subsubsection{\revisiontwo{Reflection and Generative AI}}

\revisiontwo{Following the discussion in Section~\ref{sec:discussion:generative-ai}, future work could investigate how structured reflection and generative AI interact in software engineering education: for example, the extent to which students use generative AI tools when writing reflections, whether indicator classification remains robust for AI-assisted text, and whether regular structured reflection mitigates the effects of over-reliance on such tools. Evaluations of any deployment (as above) should also consider how automated feedback on reflections interacts with students' use of generative AI, and how deployments can comply with institutional policies on AI-based tools.}

\subsubsection{Framework Refinement}

\revisiontwo{Following the reliability trade-offs discussed in Section~\ref{sec:discussion:annotation}, the} moderate agreement for Reasoning and Perspective suggests opportunities for refining these indicator definitions. Further investigation could explore whether splitting or combining certain indicators would improve both human agreement and automated classification. Additionally, examining whether all indicators are truly independent---as we currently assume---would inform whether the multi-label classification approach is optimal\revision{; co-occurrence analysis of indicators within responses would be one approach to this}. Furthermore, an investigation for elements of reflection that may exist but are not covered by our indicators that could inform further opportunities for feedback, particularly in different disciplines, or with industry practitioners.

\subsubsection{Validation Across Other Datasets}
\label{sec:discussion:future-work:validation-other-datasets}

Our classifier was developed and evaluated on a single institutional dataset\revisiontwo{, which we identified as a threat to external validity in Section~\ref{sec:discussion:limitations:external-validity}}. Future work should evaluate its performance on reflections from different institutions, countries, and student populations to establish generalisability. This includes assessing transfer performance of the classifier models presented in this study, determining whether human validation is needed before deployment in new contexts, and investigating whether fine-tuning using annotated data from the new context (i.e., reflection texts labelled with indicators) is needed to maintain performance in other settings.

\subsubsection{Generalisation to Other Contexts}

\revisiontwo{Given the differences between our dataset and those of prior work outlined in Section~\ref{sec:discussion:dataset}, evaluating} the classifier's performance on reflections from different formats (e.g., free-form essays, different prompt structures) and different educational contexts (e.g., short-term projects, industry internships) would establish the framework's broader applicability. This may require collecting additional annotated datasets or exploring transfer learning approaches, or adapting the classifier to operate at different granularities (e.g., sentence-level versus full-response classification). 

\subsubsection{Analysis of Indicator Distribution Across Questions}
\label{sec:discussion:future-work:indicator-distribution}

Whilst our three-question reflection format was designed to elicit specific elements of reflection (e.g., retrospective analysis in Question 1 and Question 2, but future planning in Question 3), we have not analysed whether associated indicators appear in the expected questions. Examining the relationship between question structure and indicator presence, in association with pedagogy of what \textit{should} appear for each question, could inform more targeted prompt design and enable question-specific feedback\revisiontwo{, directly supporting the feedback approach described in Section~\ref{sec:discussion:actionable-feedback}, where educators determine which indicators they expect for a given question}.

\subsubsection{Longitudinal Analysis}

The framework currently assesses which indicators are present in individual reflections but does not capture temporal patterns. Extending the framework to analyse temporal patterns in student reflections could provide insights into how reflection skills develop over time. This might involve tracking individual students' progression across multiple reflection submissions or identifying common developmental trajectories. Such approaches would allow for more intelligent, tailored feedback to individual students. 

\subsubsection{Application to Software Engineering Practice}
Whilst this study focused on software engineering training, the framework and classifier could potentially support reflection in professional software engineering practice\revisiontwo{, where reflection is embedded in team-based practices and directed at process improvement (see Section~\ref{sec:introduction})}. Retrospectives and post-incident reviews could benefit from automated analysis to identify missing or under-explored elements of reflection, for example as a preparation for retrospectives. Future work could investigate whether the indicators transfer to professional contexts, how they might need adapting for team-level versus individual reflections, and whether automated analysis could benefit existing agile practices. \section{Conclusion}
\label{sec:conclusion}

This study presents a reflective writing framework and automated classifier models for assessing student reflections in software engineering education. Our framework achieved inter-rater agreement comparable to or exceeding related work. The fine-tuned RoBERTa encoder-only models demonstrated strong classification performance.

We compared encoder-only and decoder-only models for classifying reflection indicators. Whilst decoder-only models such as GPT and Qwen performed reasonably well when provided with the same annotation guidelines as human annotators, the fine-tuned encoder-only models---particularly after hyperparameter tuning---achieved superior classification performance. Beyond accuracy, the computational efficiency differences were substantial: encoder-only models classified reflections in under 0.1 seconds compared to 20--45 seconds for decoder-only models. This performance gap suggests distinct use cases: encoder-only models are better suited for production deployment where real-time feedback is desired, whilst decoder-only models may be more appropriate for prototyping new frameworks or contexts where annotated training data is unavailable.

The framework provides a foundation for educators to support students' reflection skills through scalable, structured feedback in project-based software engineering courses. By automating the identification of reflection indicators, the classifier can reduce instructor workload whilst enabling more consistent and timely feedback than manual assessment alone. However, the moderate agreement on some indicators, along with the single-institution dataset, highlight areas for further refinement and validation.

\section*{\revisiontwo{Acknowledgements}}
\revisiontwo{We would like to thank the editor and the reviewers for their valuable and constructive feedback throughout the review process, which has substantially improved this paper. We also thank the students of the software engineering project course who participated in this study.}

\section*{Declarations}

\textbf{Funding:} This work did not receive external funding.
\newline
\newline
\noindent \textbf{Ethical approval:} The collection and usage of data have been approved by the University of Canterbury's Human Research Ethics Committee.
\newline
\newline
\noindent \textbf{Informed consent:} Informed consent was obtained from all individual participants included in the study.
\newline
\newline
\noindent \textbf{Author Contributions:} 
\newline
\textbf{Matthew Minish:} Conceptualisation, Methodology, Investigation, Formal analysis, Data curation, Software, Writing - original draft, Writing - review \& editing, Validation, Visualisation. 
\newline
\textbf{Matthias Galster:} Conceptualisation, Methodology, Investigation, Writing - review \& editing, Supervision. 
\newline
\textbf{Fabian Gilson:} Conceptualisation, Methodology, Investigation, Writing - review \& editing, Supervision.
\newline
\newline
\noindent \textbf{Data availability:} Our code artefacts, annotation guidelines, and other supplementary material are available publicly: \url{https://doi.org/10.5281/zenodo.15770043}
\newline
\newline
\noindent \textbf{Conflict of interest:} The authors have no competing interests to declare that are relevant to the content of this article.
\newline
\newline
\noindent \textbf{Clinical Trial Number:} Not applicable.

\bibliographystyle{spbasic}      \bibliography{my-library}

\appendix
\section{Full Model Performance Results}
\label{appendix:full-model-comparison}

Table~\ref{tab:bert-models-comparison} presents the full per-indicator results for the five encoder-only models evaluated using 10-fold cross-validation, and Table~\ref{tab:final_performance} presents the full per-indicator results for the two final RoBERTa model variants after hyperparameter tuning.

\begin{table*}[htbp]
\centering
\caption{Multi-label Text Classifier Model Performance Comparison (mean ± std)}
\label{tab:bert-models-comparison}
\resizebox{\textwidth}{!}{\begin{tabular}{lrrrrr}
\hline
\textbf{Metric} & \textbf{ALBERT} & \textbf{BERT} & \textbf{DistilBERT} & \textbf{DeBERTa} & \textbf{RoBERTa} \\
\hline
\multicolumn{6}{l}{\textbf{Overall Performance}} \\
Exact Match Accuracy & \revision{0.242±0.070} & \revision{0.283±0.023} & \revision{0.302±0.043} & \revision{0.303±0.041} & \revision{\textbf{0.325±0.042}} \\
Hamming Loss & \revision{0.184±0.038} & \revision{0.157±0.015} & \revision{0.152±0.010} & \revision{0.148±0.012} & \revision{\textbf{0.136±0.010}} \\
\revision{Macro Precision} & \revision{0.696±0.046} & \revision{0.711±0.033} & \revision{0.721±0.022} & \revision{0.733±0.030} & \revision{\textbf{0.759±0.030}} \\
\revision{Macro Recall} & \revision{0.753±0.158} & \revision{0.794±0.043} & \revision{0.793±0.033} & \revision{\textbf{0.820±0.048}} & \revision{0.817±0.054} \\
\revision{Macro F1} & \revision{0.697±0.093} & \revision{0.742±0.016} & \revision{0.750±0.018} & \revision{0.766±0.018} & \revision{\textbf{0.777±0.029}} \\
\revision{Micro Precision} & \revision{0.712±0.049} & \revision{0.739±0.037} & \revision{0.748±0.026} & \revision{0.749±0.032} & \revision{\textbf{0.776±0.033}} \\
\revision{Micro Recall} & \revision{0.770±0.166} & \revision{0.821±0.032} & \revision{0.822±0.028} & \revision{\textbf{0.842±0.040}} & \revision{0.833±0.043} \\
\revision{Micro F1} & \revision{0.726±0.099} & \revision{0.776±0.016} & \revision{0.782±0.016} & \revision{0.791±0.014} & \revision{\textbf{0.802±0.015}} \\
\hline
\multicolumn{6}{l}{\textbf{Description}} \\
Accuracy & \revision{0.763±0.029} & \revision{0.792±0.032} & \revision{0.799±0.027} & \revision{0.803±0.025} & \revision{\textbf{0.812±0.029}} \\
Precision & \revision{0.733±0.068} & \revision{0.740±0.056} & \revision{0.748±0.054} & \revision{0.754±0.039} & \revision{\textbf{0.784±0.041}} \\
Recall & \revision{0.742±0.128} & \revision{0.810±0.040} & \revision{\textbf{0.814±0.051}} & \revision{0.811±0.046} & \revision{0.781±0.049} \\
F1-Score & \revision{0.724±0.060} & \revision{0.771±0.027} & \revision{0.777±0.027} & \revision{0.780±0.024} & \revision{\textbf{0.781±0.029}} \\
\hline
\multicolumn{6}{l}{\textbf{Understanding}} \\
Accuracy & \revision{0.816±0.137} & \revision{\textbf{0.868±0.033}} & \revision{0.867±0.026} & \revision{0.866±0.033} & \revision{0.868±0.030} \\
Precision & \revision{0.896±0.043} & \revision{0.895±0.029} & \revision{0.888±0.029} & \revision{0.897±0.033} & \revision{\textbf{0.905±0.030}} \\
Recall & \revision{0.873±0.205} & \revision{0.938±0.023} & \revision{\textbf{0.946±0.029}} & \revision{0.934±0.037} & \revision{0.926±0.043} \\
F1-Score & \revision{0.863±0.145} & \revision{\textbf{0.916±0.022}} & \revision{0.916±0.018} & \revision{0.914±0.022} & \revision{0.914±0.020} \\
\hline
\multicolumn{6}{l}{\textbf{Feelings}} \\
Accuracy & \revision{0.788±0.038} & \revision{0.803±0.028} & \revision{0.817±0.037} & \revision{0.829±0.029} & \revision{\textbf{0.843±0.020}} \\
Precision & \revision{0.625±0.088} & \revision{0.644±0.073} & \revision{0.682±0.086} & \revision{0.709±0.073} & \revision{\textbf{0.744±0.079}} \\
Recall & \revision{0.729±0.123} & \revision{\textbf{0.750±0.061}} & \revision{0.724±0.087} & \revision{0.728±0.130} & \revision{0.731±0.126} \\
F1-Score & \revision{0.663±0.070} & \revision{0.688±0.036} & \revision{0.695±0.059} & \revision{0.707±0.059} & \revision{\textbf{0.724±0.059}} \\
\hline
\multicolumn{6}{l}{\textbf{Reasoning}} \\
Accuracy & \revision{0.735±0.037} & \revision{0.767±0.024} & \revision{0.763±0.038} & \revision{0.767±0.031} & \revision{\textbf{0.775±0.034}} \\
Precision & \revision{0.630±0.117} & \revision{0.649±0.066} & \revision{0.644±0.065} & \revision{0.630±0.073} & \revision{\textbf{0.655±0.084}} \\
Recall & \revision{0.689±0.188} & \revision{0.711±0.091} & \revision{0.702±0.089} & \revision{\textbf{0.784±0.075}} & \revision{0.744±0.087} \\
F1-Score & \revision{0.625±0.069} & \revision{0.671±0.031} & \revision{0.666±0.043} & \revision{\textbf{0.693±0.043}} & \revision{0.689±0.050} \\
\hline
\multicolumn{6}{l}{\textbf{Perspective}} \\
Accuracy & \revision{0.734±0.065} & \revision{0.756±0.041} & \revision{0.763±0.053} & \revision{0.757±0.066} & \revision{\textbf{0.785±0.056}} \\
Precision & \revision{0.421±0.114} & \revision{0.434±0.071} & \revision{0.447±0.087} & \revision{0.454±0.098} & \revision{\textbf{0.505±0.116}} \\
Recall & \revision{0.601±0.201} & \revision{0.593±0.124} & \revision{0.597±0.115} & \revision{\textbf{0.653±0.126}} & \revision{0.652±0.199} \\
F1-Score & \revision{0.470±0.083} & \revision{0.492±0.062} & \revision{0.505±0.087} & \revision{0.524±0.080} & \revision{\textbf{0.537±0.133}} \\
\hline
\multicolumn{6}{l}{\textbf{New Learning}} \\
Accuracy & \revision{0.852±0.042} & \revision{0.879±0.037} & \revision{0.900±0.014} & \revision{0.910±0.028} & \revision{\textbf{0.931±0.024}} \\
Precision & \revision{0.560±0.091} & \revision{0.618±0.104} & \revision{0.674±0.086} & \revision{0.720±0.134} & \revision{\textbf{0.781±0.132}} \\
Recall & \revision{0.786±0.142} & \revision{0.829±0.070} & \revision{0.820±0.066} & \revision{0.845±0.080} & \revision{\textbf{0.869±0.060}} \\
F1-Score & \revision{0.643±0.086} & \revision{0.701±0.076} & \revision{0.735±0.053} & \revision{0.764±0.054} & \revision{\textbf{0.812±0.063}} \\
\hline
\multicolumn{6}{l}{\textbf{Hindsight}} \\
Accuracy & \revision{0.941±0.030} & \revision{0.943±0.014} & \revision{0.947±0.015} & \revision{0.951±0.027} & \revision{\textbf{0.953±0.019}} \\
Precision & \revision{0.800±0.089} & \revision{0.806±0.092} & \revision{0.799±0.073} & \revision{\textbf{0.830±0.112}} & \revision{0.820±0.088} \\
Recall & \revision{0.825±0.194} & \revision{0.833±0.103} & \revision{0.858±0.080} & \revision{0.873±0.060} & \revision{\textbf{0.884±0.065}} \\
F1-Score & \revision{0.791±0.138} & \revision{0.810±0.048} & \revision{0.824±0.053} & \revision{0.846±0.073} & \revision{\textbf{0.847±0.056}} \\
\hline
\multicolumn{6}{l}{\textbf{Future Intention}} \\
Accuracy & \revision{0.898±0.088} & \revision{0.933±0.024} & \revision{0.928±0.015} & \revision{0.935±0.016} & \revision{\textbf{0.942±0.020}} \\
Precision & \revision{\textbf{0.901±0.041}} & \revision{0.900±0.062} & \revision{0.886±0.058} & \revision{0.866±0.058} & \revision{0.876±0.049} \\
Recall & \revision{0.781±0.263} & \revision{0.886±0.076} & \revision{0.885±0.064} & \revision{0.935±0.039} & \revision{\textbf{0.948±0.036}} \\
F1-Score & \revision{0.798±0.219} & \revision{0.889±0.040} & \revision{0.882±0.026} & \revision{0.897±0.027} & \revision{\textbf{0.909±0.029}} \\
\hline
\end{tabular}
}
\begin{center}
\footnotesize
\textbf{Note:} Best values for each metric are highlighted in bold.
\end{center}
\end{table*}  

\begin{table*}[htbp]
\centering
\caption{RoBERTa Model Tuning Performance Comparison (mean ± std)}
\label{tab:final_performance}
\begin{tabular}{lrrr}
\hline
\textbf{Metric} & \textbf{\makecell{RoBERTa\\(before tuning)}} & \textbf{\makecell{RoBERTa\\(F1 Optimised)}} & \textbf{\makecell{RoBERTa\\(Recall Preference)}} \\
\hline
\multicolumn{4}{l}{\textbf{Overall Performance}} \\
Exact Match Accuracy & \revision{0.325±0.042} & \revision{\textbf{0.380±0.040}} & \revision{0.285±0.052} \\
Hamming Loss & \revision{0.136±0.010} & \revision{\textbf{0.119±0.011}} & \revision{0.162±0.026} \\
\revision{Macro Precision} & \revision{0.759±0.030} & \revision{\textbf{0.774±0.024}} & \revision{0.679±0.045} \\
\revision{Macro Recall} & \revision{0.817±0.054} & \revision{0.859±0.027} & \revision{\textbf{0.915±0.021}} \\
\revision{Macro F1} & \revision{0.777±0.029} & \revision{\textbf{0.809±0.013}} & \revision{0.770±0.025} \\
\revision{Micro Precision} & \revision{0.776±0.033} & \revision{\textbf{0.789±0.024}} & \revision{0.693±0.042} \\
\revision{Micro Recall} & \revision{0.833±0.043} & \revision{0.880±0.025} & \revision{\textbf{0.929±0.018}} \\
\revision{Micro F1} & \revision{0.802±0.015} & \revision{\textbf{0.831±0.011}} & \revision{0.793±0.022} \\
\hline
\multicolumn{4}{l}{\textbf{Description}} \\
Accuracy & \revision{0.812±0.029} & \revision{\textbf{0.829±0.029}} & \revision{0.802±0.033} \\
Precision & \revision{0.784±0.041} & \revision{\textbf{0.792±0.036}} & \revision{0.710±0.057} \\
Recall & \revision{0.781±0.049} & \revision{0.827±0.087} & \revision{\textbf{0.929±0.040}} \\
F1-Score & \revision{0.781±0.029} & \revision{\textbf{0.805±0.040}} & \revision{0.802±0.026} \\
\hline
\multicolumn{4}{l}{\textbf{Understanding}} \\
Accuracy & \revision{0.868±0.030} & \revision{\textbf{0.888±0.020}} & \revision{0.888±0.022} \\
Precision & \revision{\textbf{0.905±0.030}} & \revision{0.894±0.021} & \revision{0.889±0.027} \\
Recall & \revision{0.926±0.043} & \revision{0.969±0.025} & \revision{\textbf{0.977±0.016}} \\
F1-Score & \revision{0.914±0.020} & \revision{0.930±0.014} & \revision{\textbf{0.931±0.013}} \\
\hline
\multicolumn{4}{l}{\textbf{Feelings}} \\
Accuracy & \revision{0.843±0.020} & \revision{\textbf{0.877±0.028}} & \revision{0.822±0.048} \\
Precision & \revision{0.744±0.079} & \revision{\textbf{0.758±0.077}} & \revision{0.656±0.102} \\
Recall & \revision{0.731±0.126} & \revision{0.854±0.058} & \revision{\textbf{0.884±0.057}} \\
F1-Score & \revision{0.724±0.059} & \revision{\textbf{0.800±0.051}} & \revision{0.745±0.053} \\
\hline
\multicolumn{4}{l}{\textbf{Reasoning}} \\
Accuracy & \revision{0.775±0.034} & \revision{\textbf{0.792±0.047}} & \revision{0.725±0.060} \\
Precision & \revision{0.655±0.084} & \revision{\textbf{0.661±0.092}} & \revision{0.567±0.077} \\
Recall & \revision{0.744±0.087} & \revision{0.839±0.057} & \revision{\textbf{0.902±0.052}} \\
F1-Score & \revision{0.689±0.050} & \revision{\textbf{0.732±0.043}} & \revision{0.691±0.050} \\
\hline
\multicolumn{4}{l}{\textbf{Perspective}} \\
Accuracy & \revision{0.785±0.056} & \revision{\textbf{0.823±0.032}} & \revision{0.712±0.053} \\
Precision & \revision{0.505±0.116} & \revision{\textbf{0.572±0.080}} & \revision{0.405±0.051} \\
Recall & \revision{0.652±0.199} & \revision{0.649±0.126} & \revision{\textbf{0.835±0.073}} \\
F1-Score & \revision{0.537±0.133} & \revision{\textbf{0.595±0.048}} & \revision{0.541±0.041} \\
\hline
\multicolumn{4}{l}{\textbf{New Learning}} \\
Accuracy & \revision{0.931±0.024} & \revision{\textbf{0.943±0.013}} & \revision{0.893±0.033} \\
Precision & \revision{0.781±0.132} & \revision{\textbf{0.789±0.056}} & \revision{0.638±0.084} \\
Recall & \revision{0.869±0.060} & \revision{0.912±0.050} & \revision{\textbf{0.920±0.040}} \\
F1-Score & \revision{0.812±0.063} & \revision{\textbf{0.844±0.034}} & \revision{0.749±0.049} \\
\hline
\multicolumn{4}{l}{\textbf{Hindsight}} \\
Accuracy & \revision{0.953±0.019} & \revision{\textbf{0.960±0.013}} & \revision{0.935±0.030} \\
Precision & \revision{0.820±0.088} & \revision{\textbf{0.866±0.070}} & \revision{0.742±0.120} \\
Recall & \revision{0.884±0.065} & \revision{0.869±0.050} & \revision{\textbf{0.918±0.042}} \\
F1-Score & \revision{0.847±0.056} & \revision{\textbf{0.864±0.038}} & \revision{0.813±0.070} \\
\hline
\multicolumn{4}{l}{\textbf{Future Intention}} \\
Accuracy & \revision{\textbf{0.942±0.020}} & \revision{0.937±0.013} & \revision{0.923±0.029} \\
Precision & \revision{\textbf{0.876±0.049}} & \revision{0.859±0.029} & \revision{0.826±0.063} \\
Recall & \revision{0.948±0.036} & \revision{0.955±0.023} & \revision{\textbf{0.960±0.022}} \\
F1-Score & \revision{\textbf{0.909±0.029}} & \revision{0.903±0.014} & \revision{0.886±0.037} \\
\hline
\end{tabular}
\begin{center}
\footnotesize
\textbf{Note:} Best values for each metric are highlighted in bold.
\end{center}
\end{table*}    
\end{document}